\documentclass[prd,preprintnumbers,superscriptaddress,nofootinbib]{revtex4} 

\usepackage{graphicx}
\usepackage{dcolumn}
\usepackage{bm}
\usepackage{latexsym}
\usepackage{amsfonts}
\usepackage{amssymb}
\usepackage{amsmath}
\usepackage{verbatim}
\usepackage{color}
\usepackage{natbib}
\begin{document}

\preprint{IPMU13-0064}

\title{Nonlinear stability of cosmological solutions in massive gravity}

\author{Antonio De Felice}
\email{antoniod@nu.ac.th}
\affiliation{ThEP's CRL, NEP, The Institute for Fundamental Study, Naresuan University,
Phitsanulok 65000, Thailand}
\affiliation{Thailand Center of Excellence in Physics, Ministry of Education, Bangkok 10400, Thailand}

\author{A. Emir G\"umr\"uk\c{c}\"uo\u{g}lu}
\email{emir.gumrukcuoglu@ipmu.jp}
\affiliation{Kavli Institute for the Physics and Mathematics of the Universe (WPI), Todai Institutes for Advanced Study, University of Tokyo, 5-1-5 Kashiwanoha, Kashiwa, Chiba 277-8583, Japan}

\author{Chunshan Lin}
\email{chunshan.lin@ipmu.jp}
\affiliation{Kavli Institute for the Physics and Mathematics of the Universe (WPI), Todai Institutes for Advanced Study, University of Tokyo, 5-1-5 Kashiwanoha, Kashiwa, Chiba 277-8583, Japan}

\author{Shinji Mukohyama}
\email{shinji.mukohyama@ipmu.jp}
\affiliation{Kavli Institute for the Physics and Mathematics of the Universe (WPI), Todai Institutes for Advanced Study, University of Tokyo, 5-1-5 Kashiwanoha, Kashiwa, Chiba 277-8583, Japan}

\date{\today}

\begin{abstract}
 We investigate nonlinear stability of two classes of cosmological
 solutions in massive gravity: isotropic Friedmann-Lema\^\i
 tre-Robertson-Walker (FLRW) solutions and anisotropic FLRW
 solutions. For this purpose we construct the linear cosmological
 perturbation theory around axisymmetric Bianchi type--I backgrounds. We
 then expand the background around the two classes of solutions, which
 are fixed points of the background evolution equation, and analyze
 linear perturbations on top of it. This provides a consistent
 truncation of nonlinear perturbations around these fixed point
 solutions and allows us to analyze nonlinear stability in a simple
 way. In particular, it is shown that isotropic FLRW solutions exhibit
 nonlinear ghost instability. On the other hand, anisotropic FLRW
 solutions are shown to be ghost-free for a range of parameters and
 initial conditions. 
\end{abstract}

\maketitle

\section{Introduction}

Our current understanding of gravitation is based on general relativity
(GR), whose predictions are consistent with all available experimental 
and observational data. To this end, in many cases the so called
parameterized post-Newtonian (PPN) formalism has been adopted to
quantify possible deviation from GR and to compare with experiments and 
observations. There are indeed many theoretical models that fall into
the scope of the PPN formalism. While the PPN formalism has been useful,
however, it does not include finite range modification of GR. From
phenomenological viewpoints, in analogy with the finite range nature of
the weak interaction in the standard model of particle physics, one
might argue that the possibility of finite range modification of GR
should be taken into account when experimental and observational data
are used to constrain deviation from GR. Additionally, such an extension
is also expected to alter behavior of gravity at scales larger than the
Compton wavelength of the graviton; a graviton mass as small as the
current expansion rate of the universe may provide the source for the
accelerated expansion alternative to dark energy.

From the theoretical point of view, construction of a finite-range
gravity theory has been a long-standing problem. The massive extension
of GR at the linearized level was first considered by Fierz and Pauli 
\cite{Fierz:1939ix} in 1939. It was then pointed out in 1970 that the 
linear theory suffers from a discontinuity \cite{vanDam:1970vg,
Zakharov:1970cc} in the massless limit, now known as the vDVZ
discontinuity. The discontinuity can be alleviated by allowing for
nonlinear terms \`a la Vainshtein \cite{Vainshtein:1972sx} but, as pointed
out in 1972 by Boulware and Deser \cite{Boulware:1973my}, generic
nonlinear massive theories contain an unwanted sixth degree, in
addition to the five polarizations of massive spin--2 field, which leads
to ghost instability. Since then, it took a long time until a nonlinear 
extension of the Fierz-Pauli theory without the Boulware-Deser ghost,
dubbed the dRGT theory, was found recently \cite{deRham:2010ik,
deRham:2010kj}, where the unwanted extra degree is successfully
eliminated by construction \cite{Hassan:2011hr}. However, the absence of
the sixth mode does not necessarily imply that the theory is
healthy. For instance, in some backgrounds, one (or more) of the five
polarizations of massive graviton may be superluminal
\cite{Gruzinov:2011sq, deRham:2011pt, Deser:2012qx}, or may become a
ghost \cite{Higuchi:1986py}.

The analysis of the degrees of freedom in the dRGT theory, which
originally uses the Minkowski fiducial metric, are done by considering 
the Minkowski background (see e.g.\cite{deRham:2010ik}), which resides
in the ``normal branch''. On the other hand, for the theory to have a
phenomenological relevance, it is essential to obtain non-trivial
background solutions which can describe the cosmological
evolution. However, in the normal branch no cosmological solution exists
\cite{D'Amico:2011jj}. If the theory is extended to allow for a general
fiducial metric \footnote{The dRGT theory extended in this way was shown
to be free of Boulware-Deser instability in Ref.\cite{Hassan:2011tf}.},
expanding universe solutions can be realized in the normal branch
\cite{Gumrukcuoglu:2011zh, Fasiello:2012rw, Langlois:2012hk}, but the
Higuchi bound \cite{Higuchi:1986py} cannot be respected; for instance,
for de Sitter reference metric, the scalar graviton becomes a ghost at
expansion rates higher than the graviton mass
\cite{Fasiello:2012rw}.\footnote{An exception to this conclusion may be
provided if gravity is partially massless, i.e. if there exists an
additional symmetry which forces the Higuchi bound to be saturated,
leading to the disappearance of the scalar mode within five
propagating degrees of freedom \cite{deRham:2012kf}. However, there are
indications that such a construction cannot be obtained from the dRGT
theory \cite{Deser:2013uy, deRham:2013wv}. }

In the presence of a negative spatial curvature
\cite{Gumrukcuoglu:2011ew} or a general reference metric
\cite{Gumrukcuoglu:2011zh}, there exists a second branch of solutions,
dubbed the ``self-accelerating branch'', which stems from the factorized
form of the temporal St\"uckelberg equation of motion (or equivalently, from
the Bianchi identities). With the strict Friedmann-Lema\^\i
tre-Robertson-Walker (FLRW) symmetry, this branch is disconnected from
the normal branch and as a result, the dynamics of 
perturbations are dramatically different; even though the massive
theories should have five degrees of freedom after the removal of the
Boulware-Deser degree, only two degrees (the two tensor modes in the
standard SO(3) decomposition) propagate at the linear level
\cite{Gumrukcuoglu:2011zh}. The resulting linear theory is thus very
similar to GR, with the exception of a non-vanishing and time-dependent
mass for the two gravitational wave polarizations, which may have left a
characteristic signature on observables \cite{Gumrukcuoglu:2012wt}. This
peculiar behavior may be considered as a side-effect of the FLRW
symmetry, which leads to the disconnection of the two branches. In fact,
as we show in Sec. \ref{sec:background}, the introduction of
anisotropies to the background and thus deviation from the FLRW symmetry
break the factorized form of the temporal St\"uckelberg equation,
leading to a coupling between the two branches. Around such deformed
backgrounds, all five graviton polarizations are generically dynamical
at linear order.

From the technical point of view, one of the goals of the present paper
is to construct the theory of cosmological perturbations around an
axisymmetric Bianchi type--I universe in dRGT theory. From physical
viewpoints, on the other hand, the goal of the paper is to investigate
nonlinear stability of two distinct classes of cosmological solutions:
self-accelerating isotropic FLRW solutions~\cite{Gumrukcuoglu:2011ew,
Gumrukcuoglu:2011zh} and anisotropic FLRW
solutions~\cite{Gumrukcuoglu:2012aa}. 
We will achieve this goal by
allowing for small deviation of the background from the above mentioned
two classes of solutions, applying the general formalism to analyze
perturbations around those deformed backgrounds and then considering
such linear perturbations as leading nonlinear perturbations around the
undeformed backgrounds.\footnote{This is qualitatively similar to the 
instability encountered in the self-accelerating solutions at the decoupling 
limit, arising once the vectors are turned on \cite{Koyama:2011wx}.}

The resulting quadratic kinetic action, which is
non-zero for all five degrees at linear order in background
deformations, corresponds to the cubic kinetic terms from the
perspective of undeformed backgrounds.

This approach, previously reported in \cite{DeFelice:2012mx} for
isotropic FLRW solutions, reveals that the isotropic FLRW solution
always has a ghost mode at nonlinear order with a gapless dispersion
relation. This conclusion is valid for all homogeneous and isotropic
solutions in the self-accelerating branch.

Although the deformation of the FLRW background with small anisotropy
leads to an inevitable ghost mode, there may still be healthy regions
with relatively large anisotropy. This is in analogy with scalar field
models of ghost condensation~\cite{ArkaniHamed:2003uy,
ArkaniHamed:2005gu}. As the second application of the formalism
developed in the first half of the present paper, we thus explore this 
possibility, by considering the anisotropic FLRW attractor solution of
Ref.\cite{Gumrukcuoglu:2012aa} as the undeformed background solution. On
this attractor, the expansion of the physical metric is isotropic, with
a de Sitter evolution. The fiducial metric also exhibits a de Sitter
expansion but is actually anisotropic from the viewpoint of comoving
observers residing on the physical metric. This background thus contains 
anisotropy which can only be probed by perturbations, whose evolution
depends on the alignment of the privileged direction. Applying the
general perturbation theory to this example yields that on the
attractor, there are still two degrees of freedom which do not propagate
at linear order. By adding homogeneous deformations to the background,
we find that the perturbations can in principle have a stable behavior
at nonlinear order, depending on the parameters and initial conditions. This result suggests the existence of a healthy region between the isotropic and anisotropic FLRW solutions.

The paper is organized as follows: we start by reviewing the massive
gravity action in Sec.\ref{sec:action}. Without choosing the solution,
we first present the equations for an axisymmetric Bianchi--I background
in Sec. \ref{sec:background}, and metric perturbations in
Sec.\ref{sec:perturbations}. In Sec.\ref{sec:stability}, we expand the 
background with respect to small anisotropy and discuss the implications
regarding the nonlinear stability of the isotropic FLRW solution (this
is a detailed calculation of the results already presented in 
\cite{DeFelice:2012mx}). In Sec.\ref{sec:stabilityGLM}, we consider
small deviations from the anisotropic fixed point solution of
\cite{Gumrukcuoglu:2012aa} and obtain conditions for linear and
nonlinear stability of these solutions. Finally, in
Sec.\ref{sec:discussion}, we conclude the study with a discussion of our
results. The paper is supplemented with several Appendices, where
technical details are presented.

\section{Action}
\label{sec:action}
We start this section by reviewing the action for nonlinear massive gravity in generic setup. 

The graviton mass is introduced through a non-derivative term by \cite{Boulware:1973my},
\begin{equation}
I = I_{EH,\,\Lambda}[g] + I_{\rm matter}[g,\,\psi] + I_{\rm mass}[g^{-1}f]\,,
\label{eq:action}
\end{equation}
where the first term is the Einstein-Hilbert action with a cosmological constant $\Lambda$
\begin{equation}
I_{RH,\,\Lambda}[g] = \frac{M_{Pl}}{2}\,\int d^4\,\sqrt{-g}\left(R-2\,\Lambda\right)\,,
\end{equation}
and the matter sector consists of fields $\left\{\psi\right\}$, coupled to gravity through the physical metric $g$. The existence of a massive graviton implies breaking of the general coordinate invariance, which cannot be accomplished by the physical metric $g$ alone \cite{Boulware:1973my}. For this task, the space-time tensor $f_{\mu\nu}$, dubbed the ``fiducial metric'' is introduced and parametrized as
\begin{equation}
f_{\mu\nu} \equiv \bar{f}_{AB}(\phi^C)\,\partial_\mu\phi^A\,\partial_\nu\phi^B\,,
\end{equation}
where the four scalar fields $\phi^A\,(A=0,1,2,3)$ are the St\"uckelberg fields, which are responsible for the breaking of the four gauge degrees of freedom, while $\bar{f}_{AB}(\phi^C)$ is the metric in the field space.

Imposing the absence of the BD ghost in the decoupling limit, the mass part in (\ref{eq:action}) can be constructed as \cite{deRham:2010kj}
\begin{equation}
I_{\rm mass}[g^{-1}f] = M_p^2\,m_g^2\,\int d^4 x\,\sqrt{-g}\left({\cal L}_2 + \alpha_3\,{\cal L}_3 + \alpha_4 \, {\cal L}_4\right)\,,
\end{equation}
with
\begin{eqnarray}
 {\cal L}_2 & = & \frac{1}{2}
  \left(\left[{\cal K}\right]^2-\left[{\cal K}^2\right]\right)\,, \nonumber\\
 {\cal L}_3 & = & \frac{1}{6}
  \left(\left[{\cal K}\right]^3-3\left[{\cal K}\right]\left[{\cal K}^2\right]+2\left[{\cal K}^3\right]\right), 
  \nonumber\\
 {\cal L}_4 & = & \frac{1}{24}
  \left(\left[{\cal K}\right]^4-6\left[{\cal K}\right]^2\left[{\cal K}^2\right]+3\left[{\cal K}^2\right]^2
   +8\left[{\cal K}\right]\left[{\cal K}^3\right]-6\left[{\cal K}^4\right]\right)\,,
\label{lag234}
\end{eqnarray}
where the square brackets denote the trace operation and
\begin{equation}
{\cal K}^\mu _\nu = \delta^\mu _\nu 
 - \left(\sqrt{g^{-1}f}\right)^{\mu}_{\ \nu}\,.
\label{Kdef}
\end{equation}

\section{Anisotropic background}
\label{sec:background}

In this section, we obtain the background field equations for an anisotropic extension of the FLRW physical metric.\footnote{For a more general discussion of Bianchi class A cosmologies in the context of bigravity, see \cite{Maeda:2013bha}.}
 Since one of our goals with this extension is to obtain the stability conditions of the isotropic metric, in the isotropic limit, the whole system needs to reduce to the cosmological solutions in \cite{Gumrukcuoglu:2011ew, Gumrukcuoglu:2011zh}. For this reason, we consider a fiducial metric that is of the flat FLRW form,
\begin{equation}
f_{\mu\nu} = -n^2(\phi^0) \partial_\mu \phi^0 \partial_\nu \phi^0+ \alpha^2(\phi^0) \left(\partial_\mu\phi^1\partial_\nu \phi^1+\delta_{ij}\partial_\mu\phi^i\partial_\nu \phi^j\right)\,,
\label{eq:fmn}
\end{equation}
where nonzero expectation value of the St\"uckelberg fields coincide with the space-time coordinates $\langle \phi^0 \rangle = t,\;\langle \phi^1 \rangle = x,\;\langle \phi^i \rangle = y^i$. In the rest of the paper, Greek indices span the space-time coordinates, while lower case latin indices $i,j=2,3$ correspond to the coordinates on the $y$--$z$ plane, with $y^2=y$, $y^3 = z$. 

The physical metric is chosen to be the simplest anisotropic extension of FLRW, namely, the axisymmetric Bianchi type--I metric
\begin{equation}
g^{(0)}_{\mu\nu}\,dx^\mu\,dx^\nu = - N^2(t) dt^2 + a^2(t)\,\left(e^{4\,\sigma(t)}\,dx^2 + e^{-2\,\sigma(t)}\,\delta_{ij}\,dy^i\,dy^j\right)\,.
\label{eq:g0mn}
\end{equation}

Varying the St\"uckelberg fields around the background values
\begin{equation}
\phi^A = x^A + \pi^A\,,
\label{eq:phidef}
\end{equation}
the mass term action up to first order in $\pi^a$ becomes
\begin{equation}
I_{\rm mass} = I^{(0)}_{\rm mass}+M_{Pl}^2\,m_g^2\,\int d^4x\,N\,a^3\,n\,\pi^0\,{\cal E}_\phi + {\cal O}\left[(\pi^a)^2\right]\,,
\end{equation}
giving the background equation of motion for the St\"uckelberg fields as
\begin{equation}
{\cal E}_\phi \equiv J_\phi^{(x)}\,\left(H+2\,\Sigma -H_f\,e^{-2\,\sigma}\,X\right)+2\,J_\phi^{(y)}\,\left(H-\Sigma-H_f\,e^{\sigma}\,X\right)=0\,,
\label{eq:stuckfieldeq}
\end{equation}
where 
\begin{equation}
H\equiv \frac{\dot{a}}{a\,N}\,,
\qquad
H_f\equiv \frac{\dot{\alpha}}{\alpha\,n}\,,
\qquad
\Sigma \equiv \frac{\dot{\sigma}}{N}\,,
\qquad
X \equiv \frac{\alpha}{a}\,,
\end{equation}
and
\begin{eqnarray}
J_\phi^{(x)}  &\equiv& 3- 2\,e^\sigma\,X+\alpha_3\,\left(3-4\,e^\sigma\,X+ e^{2\,\sigma}\,X^2\right) + \alpha_4\,\left(1- e^\sigma\,X\right)^2\,,\nonumber\\
J_\phi^{(y)}  &\equiv& 3- e^{-2\,\sigma}\,X-e^\sigma\,X+\alpha_3\,\left(3-2\,e^{-2\,\sigma}\,X-2\,e^\sigma\,X+ e^{-\sigma}\,X^2\right) + \alpha_4\,\left(1- e^{-2\,\sigma}\,X\right)\left(1- e^\sigma\,X\right)\,.
\label{eq:jabdef}
\end{eqnarray}
The expansion rate for the fiducial metric $H_f$ is related to the invariants of the field space metric $\bar{f}_{ab}(\phi^C)$ which are fixed by the theory, and is independent of the choice of the background values of $\phi^a$. Thus, Eq.(\ref{eq:stuckfieldeq}) can be interpreted as an algebraic equation for $\alpha$ (or equivalently for $X$), but not a differential equation. We also note that, in the isotropic limit $\left(\sigma,\,\Sigma\to0\right)$, we have $J_\phi^{(x)}=J_\phi^{(y)}$, and the corresponding equation in the FLRW case (c.f. Eq.(A.11) of Ref.~\cite{Gumrukcuoglu:2011zh}) is recovered.

Finally, we calculate the equations of motion for the metric $g_{\mu\nu}$. Since we are interested in the stability of the gravity sector only, we neglect any matter field and consider only the vacuum configuration. The independent components of the field equations for the physical metric are
\begin{eqnarray}
3\,\left(H^2 - \Sigma^2\right)-\Lambda &=& m_g^2\,\Big[-(6+4\,\alpha_3+\alpha_4)+ (3+3\,\alpha_3+\alpha_4)\,(2\,e^\sigma +e^{-2\,\sigma})\,X 
\nonumber\\
&& \qquad\qquad\qquad\qquad\qquad\qquad\qquad\;- (1+2\,\alpha_3+\alpha_4)\,(e^{2\,\sigma}+2\,e^{-\sigma})\,X^2 + (\alpha_3+\alpha_4)\,X^3\Big]\,,
\nonumber\\
3\,\left(\frac{\dot{\Sigma}}{N}+3\,H\,\Sigma\right) &=&
m_g^2 (e^{-2\,\sigma}-e^\sigma)\,X\,\Big[ (3+3\,\alpha_3+\alpha_4) - (1+2\,\alpha_3+\alpha_4)\,(e^\sigma+r)\,X+ (\alpha_3+\alpha_4)\,r\,e^\sigma\,X^2\Big]\,,\nonumber\\
-6\,\left(\frac{\dot{H}}{N}+3\,\Sigma^2 \right) &=& m_g^2\,X\,\Bigg[ (3+3\,\alpha_3+\alpha_4)\,(2\,e^\sigma+e^{-2\,\sigma}-3\,r)+(\alpha_3+\alpha_4)\,\Big[3-(e^{2\,\sigma}+2\,e^{-\sigma})\,r\Big]\,X^2\nonumber\\
&&\qquad\qquad\qquad\qquad\qquad\qquad\;+2\,(1+2\,\alpha_3+\alpha_4)\Big[(2\,e^{\sigma}+e^{-2\,\sigma})\,r-(e^{2\,\sigma}+2\,e^{-\sigma})\Big]\,X
\Bigg]\,,
\end{eqnarray}
where we have introduced
\begin{equation}
r\equiv \frac{n}{NX}\,.
\end{equation}

\section{Perturbations}
\label{sec:perturbations}

In this section, we calculate the action quadratic in perturbations around the metric (\ref{eq:g0mn}). The most general set of perturbations around the axisymmetric Bianchi type--I are given by \cite{Gumrukcuoglu:2007bx}
\begin{equation}
g^{(1)}_{\mu \nu} = \left( \begin{array}{lrrr}
- 2\,N^2\,\Phi\quad & \quad\quad a\,e^{2\,\sigma}\,N\, \partial_x \chi & a\,e^{-\sigma}\,N \left( \partial_i B + v_i \right) \\
& a^2\,e^{4\,\sigma} \psi & a^2\, e^\sigma\, \partial_x \left( \partial_i \beta + \lambda_i \right) \\
& & a^2\,e^{-2\,\sigma} \left[ \tau\,\delta_{ij} + \partial_i\partial_j E + \partial_{(i}h_{j)}\right] 
\end{array} \right) \,,
\label{eq:perts-def1}
\end{equation}
where $\partial_{(i}h_{j)} \equiv (\partial_i h_j + \partial_i h_j)/2$
and $\partial^i v_i = \partial^i \lambda_i = \partial^i h_i =0$. Note
that, since the $y$--$z$ plane is Euclidean, the indices $i,j$ are
raised and lowered with $\delta^{ij}$ and $\delta_{ij}$. Similarly, we
decompose the perturbations of the St\"uckelberg fields
(\ref{eq:phidef}) as 
\begin{equation}
\pi^A = \left(
\pi^0 ,\; \partial_1 \pi^1,\; \partial^i\pi+\pi^i
\right)\,,
\end{equation}
where $\partial_i \pi^i =0$. Since the vectors are defined on the 2d $y$--$z$ plane, the transversity condition can be used to reduce each of these vectors to a single degree of freedom 
\begin{equation}
v_i = \epsilon_i^{\;\,j}\partial_j\,v \,,
\qquad
\lambda_i = \epsilon_i^{\;\,j}\partial_j\,\lambda \,,
\qquad
h_i = \epsilon_i^{\;\,j}\partial_j\,h \,,
\qquad
\pi_i = \epsilon_i^{\;\,j}\partial_j\,\pi_{\rm odd} \,,
\end{equation}
where $\epsilon_{ij}$ is a unit anti-symmetric tensor with
$\epsilon_{23}=-\epsilon_{32}=1$ and
$\epsilon_i^{\;\,j}=\epsilon_{ik}\delta^{kj}$.

Although the nonzero $\sigma$ in the background metric (\ref{eq:g0mn}) breaks the isotropy, there is still a residual symmetry corresponding to rotations around the $x$ axis. Under such a rotation, the perturbations which transform as 2d scalars, or {\it even modes} ($\Phi$, $\chi$, $B$, $\psi$, $\beta$, $\tau$, $E$) and the perturbations which transform as 2d vectors, or {\it odd modes} ($v$, $\lambda$, $h$ ) decouple at the level of the quadratic action. This allows us to study each sector separately in the following.

\subsection{Gauge invariant variables}
Before moving on to the discussion of the action, let us first determine the transformation properties of the perturbations and build gauge invariant variables. We consider infinitesimal coordinate transformations
\begin{equation}
x^\mu \to x^\mu + \xi^\mu\,,
\end{equation}
with
\begin{equation}
\xi^A = \left(
\xi^0 ,\; \partial_1 \xi^1,\; \partial^i\xi+\epsilon^{ij}\partial_j\xi_{\rm odd}
\right)\,.
\end{equation}
Under this transformation, the even perturbations transform as
\begin{eqnarray}
\Phi &\to&\Phi + \frac{1}{N}\,\partial_t\left(N\,\xi^0\right)\,,\nonumber\\
\chi &\to& \chi - \frac{N\,e^{-2\,\sigma}}{a}\,\xi^0 + \frac{a\,e^{2\,\sigma}}{N}\,\dot{\xi}^1\,,\nonumber\\
B &\to& B - \frac{N\,e^\sigma}{a}\,\xi^0+\frac{a\,e^{-\sigma}}{N}\,\dot{\xi}\,,\nonumber\\
\psi &\to& \psi +2\,N\,\left(H+2\,\Sigma\right)\,\xi^0 + 2\,\partial_x^2\,\xi^1\,,\nonumber\\
\beta &\to& \beta + e^{3\,\sigma}\,\xi^1 + e^{-3\,\sigma}\,\xi\,,\nonumber\\
\tau &\to& \tau +2\,N\,\left(H-\Sigma\right)\,\xi^0\,,\nonumber\\
E &\to& E +2\,\xi\,,\nonumber\\
\pi^0 &\to& \pi^0 + \xi^0\,,\nonumber\\
\pi^1 &\to& \pi^1 + \xi^1\,,\nonumber\\
\pi &\to& \pi + \xi\,,
\end{eqnarray}
while for the odd perturbations, we have
\begin{eqnarray}
v &\to& v + \frac{a\,e^{-\sigma}}{N}\,\dot{\xi}_{\rm odd}\,,\nonumber\\
\lambda &\to& \lambda + e^{-3\,\sigma}\,\xi_{\rm odd}\,,\nonumber\\
h &\to& h +2\,\xi_{\rm odd}\,,\nonumber\\
\pi_{\rm odd} &\to& \pi_{\rm odd} + \xi_{\rm odd}\,.
\end{eqnarray}
Using these transformations, we first define gauge invariant variables which do not refer to the St\"uckelberg fields. For the even perturbations, we introduce
\begin{eqnarray}
\hat{\Phi} &=& \Phi- \frac{1}{2\,N}\,\partial_t\left(\frac{\tau}{H-\Sigma}\right)\,,\nonumber\\
\hat{\chi} &=& \chi + \frac{e^{-2\,\sigma}}{2\,a\,(H-\Sigma)}\tau- \frac{a\,e^{2\,\sigma}}{N}\,\partial_t \left[ e^{-3\,\sigma}\,\left(\beta - \frac{e^{-3\,\sigma}}{2}\,E\right)\right]\,,\nonumber\\
\hat{B} &=& B + \frac{e^\sigma}{2\,a\,(H-\Sigma)}\,\tau - \frac{a\,e^{-\sigma}}{2\,N}\,\dot{E}\,,\nonumber\\
\hat{\psi} &=& \psi -\frac{H+2\,\Sigma}{H-\Sigma}\,\tau - e^{-3\,\sigma}\,\partial_x^2\left(2\,\beta - e^{-3\,\sigma}\,E\right)\,,
\label{eq:gi-gr-even}
\end{eqnarray}
while for the odd perturbations, we have
\begin{eqnarray}
\hat{v} &=& v - \frac{a\,e^{-\,\sigma}}{2\,N}\,\dot{h}\,,\nonumber\\
\hat{\lambda} &=& \lambda - \frac{e^{-3\,\sigma}}{2}\,h\,.
\label{eq:gi-gr-odd}
\end{eqnarray}
The variables defined above are not sufficient to account for all of the physical degrees of freedom. Using the St\"uckelberg perturbations, we can define three even variables
\begin{eqnarray}
\hat{\tau}_\pi &=& \pi^0 - \frac{\tau}{2\,N\,(H-\Sigma)}\,,\nonumber\\
\hat{\beta}_\pi &=& \pi^1 - e^{-3\,\sigma}\,\left(\beta - \frac{e^{-3\,\sigma}}{2}\,E\right)\,,\nonumber\\
\hat{E}_\pi &=& \pi -\frac{1}{2}\,E\,, 
\label{eq:gi-st-even}
\end{eqnarray}
and one odd variable
\begin{equation}
\hat{h}_\pi = \pi_{\rm odd} - \frac{1}{2}\,h\,.
\label{eq:gi-st-odd}
\end{equation}
The gauge invariant variables defined in Eqs.(\ref{eq:gi-gr-even}) and (\ref{eq:gi-gr-odd}) originate only from the perturbations of the physical metric, thus are already present in the GR case~\cite{Himmetoglu:2008hx}. The additional degrees in Eqs.(\ref{eq:gi-st-even}) and (\ref{eq:gi-st-odd}) are the physical degrees of freedom associated with the additional degrees of the massive graviton and arise from the breaking of general coordinate invariance by the nonzero expectation value of the St\"uckelberg fields.

\subsection{Odd perturbations}
We start by the action for the odd perturbations only. The perturbed metric we consider is, from (\ref{eq:g0mn}) and (\ref{eq:perts-def1}),
\begin{equation}
ds^2_{\rm odd} =-N^2\,dt^2 +2\,a\,e^{-\sigma}\,N\,v_i\,dt\,dy^i+a^2\,e^{4\,\sigma} \,dx^2 +2\,a^2\,e^\sigma\,\partial_x\lambda_i \,dx\,dy^i+a^2\,e^{-2\,\sigma}\left(\delta_{ij}+\partial_{(i}h_{j)}\right)dy^idy^j\,,
\label{eq:oddmetric}
\end{equation}
while for the St\"uckelberg fields, we consider
\begin{equation}
\phi^0=t\,,\qquad
\phi^1=x\,,\qquad
\phi^i=y^i + \pi^i\,.
\label{eq:oddstuck}
\end{equation}
After using these decompositions in the action (\ref{eq:action}), then switching to the gauge invariant variables defined in Eqs.~(\ref{eq:gi-gr-odd}) and (\ref{eq:gi-st-odd}), the resulting action depends on the three perturbations ($\hat{v}$, $\hat{\lambda}$, $\hat{h}_\pi$). Among these, $\hat{v}$ does not have any time derivatives and can be removed by solving the constraint equation. In general relativity, this operation also removes $\hat{h}_\pi$ and the final action can be written in terms of $\hat{\lambda}$ only. In the nonlinear theory of massive gravity however, we expect that $\hat{h}_\pi$ remains in the action as an extra degree of freedom coming from the St\"uckelberg sector. 

We Fourier transform perturbations as
\begin{equation}
\delta (t,\,x,\,y^i) = \int \frac{dk_L\,dk_T^2}{(2\,\pi)^{3/2}}\,e^{i\,(k_L\,x+k_{i}\,y^i)}\,\delta(t,\,k_L\,,k_{i})\,,
\end{equation}
where $k_L$ and $k_i$ are the components of the comoving momentum in the $\hat{x}$ and $\hat{y}^i$ directions, respectively. Due to the 2d rotational symmetry around the $\hat{x}$ axis, the physical results are expected to depend only on the longitudinal component $k_L$ and the magnitude of the transverse components $k_T \equiv \sqrt{k_{2}^2+k_{3}^2}$.

Since the quadratic action does not depend on the derivatives of $\hat{v}$, its equation of motion can be solved by
\begin{equation} 
\hat{v} = \frac{a\,e^\sigma}{N\,\left[p^2(e^\sigma+r)+2\,e^{2\,\sigma}\,m_g^2\,X\,J_\phi^{(y)} \right]}\,\left[2\,m_g^2\,X\,J_\phi^{(y)} \,\dot{\hat{h}}_\pi
+e^{-2\sigma}\,p_L^2\,\left(e^\sigma+r\right)\,\frac{d}{dt}\left(e^{3\,\sigma}\,\hat{\lambda}\right)
\right]\,,
\end{equation}
where $J_\phi^{(y)}$ is defined in Eq.~(\ref{eq:jabdef}), while 
\begin{equation}
p_L \equiv \frac{k_L}{a\,e^{2\,\sigma}}\,,
\qquad
p_T \equiv \frac{k_T}{a\,e^{-\sigma}}\,,
\qquad
p^2 \equiv p_L^2 + p_T^2\,
\end{equation}
are the physical momenta in longitudinal and transverse directions, and
the total physical momentum, respectively. 

After a further field redefinition,
\begin{equation}
{\cal V}_1 \equiv -e^{3\,\sigma}\,\hat{\lambda} \,,
\qquad
{\cal V}_2 \equiv \frac{2\,e^{3\,\sigma}\,p_L^2}{p^2}\,\hat{\lambda} - 2\,\hat{h}_\pi\,,
\end{equation}
the quadratic action takes the following form
\begin{equation}
I^{(2)}_{\rm odd} = \frac{M_p^2}{2}\int N\,dt\,dk_L\,d^2 k_T\,a^3\,\left[K_{11}\, \frac{\left\vert\dot{\cal V}_1\right\vert^2}{N^2} + K_{22}\,\left\vert\frac{\dot{\cal V}_2}{N} - \gamma\,{\cal V}_1\right\vert^2 - \Omega^2_{11}\,\left\vert{\cal V}_1\right\vert^2- \Omega^2_{22}\,\left\vert{\cal V}_2\right\vert^2- \Omega^2_{12}\,\left({\cal V}_1^\star{\cal V}_2+{\cal V}_2^\star{\cal V}_1\right)\right]\,,
\label{eq:oddaction}
\end{equation}
where
\begin{eqnarray}
K_{11} &=& a^4\,e^{-4\,\sigma}\,\frac{p_L^2\,p_T^4}{2\,p^2}\,,\nonumber\\
K_{22} &=& \frac{m_g^2\,a^4\,e^{-2\,\sigma}\,X\,J_\phi^{(y)}\,p_T^2\,p^2}{4\,\left[2\,m_g^2\,e^{2\,\sigma}X\,J_\phi^{(y)}+ (e^\sigma+r)\,p^2\right]}\,,\nonumber\\
\gamma &=& \frac{12\,\Sigma\,p_L^2\,p^2_T}{p^4}\,,\nonumber\\
\Omega^2_{11} &=& \frac{a^4\,e^{-4\,\sigma}\,p_L^2\,p_T^4}{2}\left[1+ \frac{m_g^2\,e^\sigma\,X}{p^4}\left(J_1\,p_L^2 + \frac{2\,J_2\,p_T^2}{1+e^{-3\,\sigma}}\right)\right]\,,\nonumber\\
\Omega^2_{12} &=& \frac{m_g^2\,a^4\,X\,p_L^2\,p_T^4}{4\,p^2\,(1+e^{3\,\sigma})}\left[(1+e^{-3\,\sigma})\,J_1 -2\,J_2\right]\,,\nonumber\\
\Omega^2_{22} &=& \frac{m_g^2\,a^4\,X\,p_T^2}{8\,(1+e^{3\,\sigma})}\,\left[ 2\,J_2\,p_L^2 + (1+e^{-3\,\sigma})\,J_1\,p_T^2\right]\,,
\label{eq:oddcoeffs}
\end{eqnarray}
and
\begin{eqnarray}
J_1 &\equiv& (3+3\,\alpha_3+\alpha_4) - (1+2\,\alpha_3+\alpha_4)\,(r+e^{-2\,\sigma})\,X + (\alpha_3+\alpha_4)\,r\,e^{-2\,\sigma}\,X^2\,,\nonumber\\
J_2 &\equiv& (3+3\,\alpha_3+\alpha_4) - (1+2\,\alpha_3+\alpha_4)\,(r+e^{\sigma})\,X + (\alpha_3+\alpha_4)\,r\,e^{\sigma}\,X^2\,.
\label{eq:j12def}
\end{eqnarray}
\subsection{Even perturbations}

The perturbed metric for the even sector is, from (\ref{eq:g0mn}) and (\ref{eq:perts-def1}),
\begin{eqnarray}
ds_{\rm even}^2 &=& -N^2\left(1+2\,\Phi\right)\,dt^2 + 2\,a\,e^{2\,\sigma}\,N\,\partial_x \chi \,dt\,dx+ 2\,a\,e^{-\sigma}\,N\,\partial_i B \,dt\,dy^i
\nonumber\\
&&+ a^2\,e^{4\,\sigma}\,\left(1+\psi\right)dx^2 
+2\,a^2\,e^\sigma\,\partial_x\partial_i\beta\,dx\,dy^i + a^2\,e^{-2\,\sigma}\,\left[\delta_{ij}\left(1+\tau\right) +\partial_i\partial_j E\right]dy^i\,dy^j\,,
\end{eqnarray}
while the even perturbations of St\"uckelberg fields read
\begin{equation}
\phi^0 = t + \pi^0\,,\qquad
\phi^1 = x + \partial_x\pi^1\,,\qquad
\phi^i = y^i + \partial^i \pi\,.
\end{equation}
Applying these decompositions to the action (\ref{eq:action}), we then switch to the gauge invariant variables defined in Eqs.~(\ref{eq:gi-gr-even}) and (\ref{eq:gi-st-even}). The resulting action is then manifestly gauge invariant and depends on seven perturbations ($\hat{\Phi}$, $\hat{B}$, $\hat{\chi}$, $\hat{\psi}$, $\hat{\tau}_\pi$, $\hat{\beta}_\pi$, $\hat{E}_\pi$). 
The three degrees arising from the $g^{(1)}_{0\mu}$ perturbations, namely $\hat{\Phi}$, $\hat{B}$ and $\hat{\chi}$, are non-dynamical, thus can be integrated out. Furthermore, the kinetic term for the $\hat{\tau}_\pi$ is proportional to background equations of motion, so this degree is also non-dynamical. We interpret this as the would-be BD ghost, which is eliminated in this theory by construction. In the massless theory (GR), using the constraint equations also removes the degrees $\hat{\beta}_\pi$, $\hat{E}_\pi$, leaving only $\hat{\psi}$ in the action, which becomes one of the gravity wave polarizations in the isotropic limit. However, due to the nonzero mass of the graviton, these two degrees are generically dynamical. Thus, the metric perturbations in vacuum has three physical even perturbations.

Upon expanding the fields in momentum space, the resulting action is formally
\begin{equation}
I^{(2)}_{\rm even} = \frac{M_p^2}{2}\,\int N\,dt\,dk_L\,d^2k_T\,a^3 \left[\frac{\dot{{\cal Y}}^\dagger}{N} \,K\, \frac{\dot{{\cal Y}}}{N} -{\cal Y}^\dagger\, \Omega^2\,{\cal Y} + {\cal Z}^\dagger \,{\cal A}\,{\cal Y} + {\cal Y}^\dagger \,{\cal A}^T\,{\cal Z} +
{\cal Z}^\dagger \,{\cal B}\,\frac{\dot{\cal Y}}{N}+ \frac{\dot{\cal Y}^\dagger}{N}\,{\cal B}^T\,{\cal Z} + {\cal Z}^\dagger\,{\cal C}\,{\cal Z}
\right]\,,
\label{eq:acteven1}
\end{equation}
with explicit expression for the matrices $K$, $\Omega^2$, ${\cal A}$,
${\cal B}$ and ${\cal C}$ given in Appendix \ref{sec:evenact}. In the
above action, the basis 
${\cal Y}=(\hat{\psi},\hat{\beta},\hat{E}_\pi)^T$ corresponds to
the three dynamical degrees, while 
${\cal Z}=(\hat{\Phi}, \hat{B}, \hat{\chi}, \hat{\tau}_\pi)^T$ contains 
the four non-dynamical ones. The equation of motion for the latter can be
solved to give 
\begin{equation}
{\cal Z}= - {\cal C}^{-1}\,\left({\cal A}\,{\cal Y} + {\cal B}\,\frac{\dot{\cal Y}}{N}\right)\,,
\end{equation}
where we assumed $\det {\cal C}\neq0$ \footnote{As we show in Sec.\ref{sec:stabilityGLM}, this is a relevant assumption for the cases we study.}. 
Using this solution in (\ref{eq:acteven1}), we obtain
\begin{equation}
I^{(2)}_{\rm even} = \frac{M_p^2}{2}\,\int N\,dt\,dk_L\,d^2k_T\,a^3 \left[\frac{\dot{{\cal Y}}^\dagger}{N} \,\bar{K}\, \frac{\dot{{\cal Y}}}{N} +\frac{\dot{{\cal Y}}^\dagger}{N}\,\bar{M}\,{\cal Y}+{\cal Y}^\dagger\,\bar{M}^T\,\frac{\dot{{\cal Y}}}{N}
-{\cal Y}^\dagger\, \bar{\Omega}^2\,{\cal Y} 
\right]\,,
\label{eq:acteven2}
\end{equation}
where
\begin{equation}
\bar{K} = K -{\cal B}^T\,{\cal C}^{-1}\,{\cal B}\,,\qquad
\bar{M} = - {\cal B}^T\,{\cal C}^{-1}\,{\cal A}\,,\qquad
\bar{\Omega}^2 = \Omega^2 +{\cal A}^T\,{\cal C}^{-1}\,{\cal A}\,.
\label{eq:defKMO}
\end{equation}

\section{Nonlinear instability of FLRW solution}
\label{sec:stability}
In this section, we analyze the action for the perturbations studied in the previous section, by considering small anisotropic departure from a FLRW metric. As discussed in \cite{DeFelice:2012mx}, quadratic action in Bianchi-I background with small anisotropy, gives information on the nonlinear action in FLRW background.

Assuming $|\sigma|\ll 1$ and $|\Sigma/H|\ll 1$, we introduce the parameter $\epsilon\ll1$ and expand the quadratic action with respect to $\epsilon$. 

The St\"uckelberg equation of motion at leading order gives,
\begin{equation}
3+3\,\alpha_3+\alpha_4 - 2\,(1+2\,\alpha_3+\alpha_4)\,X + (\alpha_3+\alpha_4)\,X^2={\cal O}(\epsilon)\,,
\end{equation}
implying
\begin{eqnarray}
J_\phi^{(x)} &=& -2\,X\,\left[(1+2\,\alpha_3+\alpha_4)-(\alpha_3+\alpha_4)\,X\right]\,\sigma+{\cal O}(\epsilon^2)\,,\nonumber\\
J_\phi^{(y)} &=& X\,\left[(1+2\,\alpha_3+\alpha_4)-(\alpha_3+\alpha_4)\,X\right]\,\sigma+{\cal O}(\epsilon^2)\,,\nonumber\\
J_1 = J_2 +{\cal O}(\epsilon)&=& (1-r)\,X\,\left[(1+2\,\alpha_3+\alpha_4)-(\alpha_3+\alpha_4)\,X\right]+{\cal O}(\epsilon)\,.
\end{eqnarray}

\subsection{Odd perturbations}
Expanding the action quadratic in odd modes, the terms (\ref{eq:oddcoeffs}) at leading order in the small anisotropy expansion are obtained as
\begin{eqnarray}
K_{11} &=& \frac{a^4\,p_L^2\,p_T^4}{2\,p^2} + {\cal O}(\epsilon)\,,\nonumber\\
K_{22} &=& \frac{a^4\,p_T^2\,M_{GW}^2}{4\,(1-r^2)}\,\sigma+ {\cal O}(\epsilon^2)\,,\nonumber\\
\gamma &=& {\cal O}(\epsilon)\,,\nonumber\\
\Omega^2_{11} &=& K_{11}\,\times \left(p^2+M^2_{GW}\right) + {\cal O}(\epsilon)\,,\nonumber\\
\Omega^2_{12} &=& {\cal O}(\epsilon)\,,\nonumber\\
\Omega^2_{22} &=& K_{22}\,\times\left(c_s^2\,p^2\right)+{\cal O}(\epsilon)\,,
\end{eqnarray}
where
\begin{equation}
M_{GW}^2 = m_g^2\,(1-r)\,X^2\left[(1+2\,\alpha_3+\alpha_4)-(\alpha_3+\alpha_4)\,X\right]\,,\qquad
c_s^2 = \frac{1-r^2}{2\,\sigma}\,.
\end{equation}
Thus, at leading order, we identify the mode ${\cal V}_1$ with one of the gravity wave (GW) polarizations in FLRW background \cite{Gumrukcuoglu:2011zh}. The extra degree of freedom ${\cal V}_2$ is massless and has sound speed $c_s$. We also note that the leading order kinetic term for ${\cal V}_2$ may in general become negative, leading to a ghost degree. For $|\sigma|\ll1$, in order to avoid the ghost and to make $M_{GW}^2$ positive, we require
\begin{equation}
(1-r)\,\sigma >0\,,
\label{eq:noghost}
\end{equation}
which also agrees with the condition for avoiding the gradient instability of ${\cal V}_2$, i.e. $c_s^2 > 0$.

\subsection{Even perturbations}
\label{sec:evenexpand}

For the even sector, the components of the kinetic matrix defined in (\ref{eq:defKMO}) at leading order are
\begin{eqnarray}
\bar{K}_{11} &=& \frac{p_T^4}{8\,p^4} + {\cal O}(\epsilon)\,,\nonumber\\
\bar{K}_{12} &=& -\frac{a^2\,M_{GW}^2\,p_L^2\,(2\,p^2+p_T^2)}{2\,p^4\,(1-r^2)}\,\sigma + {\cal O}(\epsilon^2)\,,\nonumber\\
\bar{K}_{13} &=& -\frac{a^2\,M_{GW}^2\,p_T^2\,(p^2+p_L^2)}{4\,p^4\,(1-r^2)}\,\sigma + {\cal O}(\epsilon^2)\,,\nonumber\\
\bar{K}_{22} &=& - \frac{2\,a^4\,M_{GW}^2\,p_L^2}{1-r^2}\,\sigma + {\cal O}(\epsilon^2)\,,\nonumber\\
\bar{K}_{23} &=& \frac{a^4\,M_{GW}^2\,p_L^2 \,p_T^2\,\left(p^4\,r^2-3\,H^2\,M_{GW}^2\right)}{3\,H^2\,p^4\,(1-r^2)^2}\,\sigma^2+{\cal O}(\epsilon^3)\,,\nonumber\\
\bar{K}_{33} &=& \frac{a^4\,M_{GW}^2\,p_T^2}{1-r^2}\,\sigma +{\cal O}(\epsilon^2)\,.
\end{eqnarray}
Notice that at order $\epsilon$, the first mode has kinetic mixing with the second and third ones. One can rotate the basis to make the kinetic terms diagonal. The leading order terms of the eigenvalues of the matrix $\bar{K}$ can be found to be:
\begin{eqnarray}
\kappa_1 &=& \frac{p_T^4}{8\,p^4}+{\cal O}(\epsilon)\,,\nonumber\\
\kappa_2 &=& - \frac{2\,a^4\,M_{GW}^2\,p_L^2}{1-r^2}\,\sigma + {\cal O}(\epsilon^2)\,,\nonumber\\
\kappa_3 &=& \frac{a^4\,M_{GW}^2\,p_T^2}{1-r^2}\,\sigma + {\cal O}(\epsilon^2)\,.
\end{eqnarray}
The only eigenvalue which is non zero in the FLRW limit is the first one, and corresponds to one of the gravity wave polarizations. The remaining two are thus the extra degrees arising from the graviton mass. However, at leading order, the kinetic terms of these two modes are related through
\begin{equation}
\kappa_3 = -\frac{p_T^2}{2\,p_L^2}\,\kappa_2\,,
\end{equation}
which implies that for $|\sigma|\ll 1$, the two extra modes cannot
simultaneously have positive kinetic terms; one of them is always a ghost. 

The dispersion relation for each mode can be obtained by a series of transformations, as described in detail in Appendix \ref{app:diagonalize}. Assuming that the condition (\ref{eq:noghost}) is satisfied, the dispersion relations are found to be:
\begin{eqnarray}
\omega_1^2 &=& p^2+ M_{GW}^2 + {\cal O}(\epsilon) \,,\nonumber\\
\omega_2^2 &=& -\left(\frac{1-r^2}{24\,\sigma}\right)\,\left[\sqrt{\left(10\,p^2+p_T^2\right)^2 - 8\,p_L^2\,p_T^2} - \left(2\,p^2 +3\,p_T^2\right) \right]\,,\nonumber\\
\omega_3^2 &=& \left(\frac{1-r^2}{24\,\sigma}\right)\,\left[\sqrt{\left(10\,p^2+p_T^2\right)^2 - 8\,p_L^2\,p_T^2} + \left(2\,p^2 +3\,p_T^2\right) \right]\,.
\end{eqnarray}
The first mode corresponds to one of the GW polarizations in the FLRW
limit, while the other two are extra degrees of freedom. Note that in
this case, the second degree is a ghost. We also stress that dispersion
relation for the extra degrees consists only of momentum terms and thus
is gapless.

\section{Nonlinear stability of anisotropic fixed point solution}
\label{sec:stabilityGLM}
In this section, specializing the field space metric to be pure de Sitter
($H_f={\rm const.}$), we analyze the action for the perturbations studied in 
Sec.\ref{sec:perturbations}, by considering small deviation from the
anisotropic fixed point found in \cite{Gumrukcuoglu:2012aa} 
\footnote{The global stability of this fixed point solution, shown in \cite{Gumrukcuoglu:2012aa}, can be interpreted as the validity of the cosmic no-hair conjecture \cite{Wald:1983ky} in massive gravity. This point was addressed in the context of bigravity in \cite{Sakakihara:2012iq}.}
\begin{equation}
H_0,\,\sigma_0,\,X_0  {\rm\  constants} \,, 
\qquad
r_0 = e^{-2\,\sigma_0}\,.
\end{equation}

\subsection{Expanding the background for small anisotropy}
Assuming $|\sigma-\sigma_0|\ll 1$ and $|\Sigma/H_0|\ll 1$, we expand the background quantities
\begin{eqnarray}
\sigma &=& \sigma_0 + \epsilon \, \sigma_1 +{\cal O}(\epsilon^2)\,,\nonumber\\
\Sigma &=& \epsilon \, \Sigma_1 +{\cal O}(\epsilon^2)\,,\nonumber\\
H &=& H_0 + \epsilon H_1 + {\cal O}(\epsilon^2)\,,\nonumber\\
X &=& X_0 + \epsilon X_1 + {\cal O}(\epsilon^2)\,,\nonumber\\
r &=& r_0 + \epsilon r_1 + {\cal O}(\epsilon^2)\,.
\end{eqnarray}
Note that $H_f$ is not expanded since the fiducial metric is set to be
de Sitter. Even if the fiducial metric were of
general FLRW type, $H_f$ would not be expanded as long as the unitary
gauge is employed at the background level.

Let us study the background equations order by order. At order ${\cal O}(\epsilon^0)$, upon using the characteristic relation for the anisotropic fixed point, $r_0 = e^{-2\,\sigma_0}$, or,
\begin{equation}
X_0 = \frac{H_0}{H_f}\,e^{2\,\sigma_0}\,,
\end{equation}
there are two more equations at this order;
\begin{eqnarray}
H_0^2\,e^{3\,\sigma_0}(\alpha_3+\alpha_4) - H_0 H_f (1+  e^{3\,\sigma_0})(1+2\,\alpha_3+\alpha_4)+H_f^2 (3+3\,\alpha_3+\alpha_4) = 0
\,,
\nonumber\\
H_0^2 \left[3+\frac{m_g^2}{H_f^2}\,e^{3\,\sigma_0}(1+2\,\alpha_3+\alpha_4)\right] - \frac{H_0 \,m_g^2}{H_f}\,(1+e^{3\,\sigma_0})(3+3\,\alpha_3+\alpha_4)+m_g^2\,(6+4\,\alpha_3+\alpha_4)+\Lambda=0\,,
\end{eqnarray}
the solutions of which, determine the sets of fixed point values for $H_0$ and $\sigma_0$. However, noting that $\alpha_3$ and $\alpha_4$ enter these equations linearly, we can obtain a single set of solutions. Furthermore, introducing two mass scales as
\begin{eqnarray}
M^2 &\equiv& \frac{H_0\,m_g^2}{3\,H_f^3}\,\left[H_0^2 e^{3\,\sigma_0}(1+2\,e^{3\,\sigma_0})(\alpha_3+\alpha_4) - 2\,H_0\,H_f\,(1+e^{3\,\sigma_0}+e^{6\,\sigma_0})(1+2\,\alpha_3+\alpha_4) \right.\nonumber\\
&&\qquad\qquad\left.+ H_f^2(2+e^{3\,\sigma_0})(3+3\,\alpha_3+\alpha_4)\right]\,,\nonumber\\
\tilde{M}^2 &\equiv &-\frac{3\,H_0\,m_g^2}{2\,H_f^3}\,
\left[H_0^2 e^{6\,\sigma_0}(\alpha_3+\alpha_4) - 2\,H_0\,H_f\,e^{3\,\sigma_0}(1+2\,\alpha_3+\alpha_4) + H_f^2(3+3\,\alpha_3+\alpha_4)\right]\,,
\end{eqnarray}
which, along with the $\cal{O}(\epsilon^0)$ order background equations, allow us to express all physical background quantity in terms of $H_0$, $\sigma_0$, $M$ and $\tilde{M}$, in mass units of $H_f$. Thus, from here on, we choose 
 the $H_f=1$ units.

Upon using the zero order relations above, the ${\cal O}(\epsilon)$ equations yield,
\begin{eqnarray}
X_1 &=& \frac{H_0 e^{2\,\sigma_0}}{\tilde{M}^4-27\,H_0^2(3\,M^2-\tilde{M}^2)}\left\{
\left[2\,\tilde{M}^4+27\,H_0^2 (3\,M^2+2\,\tilde{M}^2)\right]\sigma_1 + 18\,H_0\,\tilde{M}^2\,\Sigma_1\right\}
\,,\nonumber\\
H_1 &=& -\frac{\tilde{M}^2}{\tilde{M}^4-27\,H_0^2(3\,M^2-\tilde{M}^2)}\left(27\,H_0\,M^2\,\sigma_1+2\,\tilde{M}^2\,\Sigma_1\right)
\,,\nonumber\\
r_1e^{2\sigma_0} &=& 
 \frac{243(3M^2+2\tilde{M}^2)(3M^2-\tilde{M}^2)H_0^4+9\tilde{M}^2H_0^2(81M^4-66M^2\tilde{M}^2+4\tilde{M}^4)-\tilde{M}^6(27M^2-2\tilde{M}^2)}{[\tilde{M}^4-27\,H_0^2(3\,M^2-\tilde{M}^2)][\tilde{M}^4+9H_0^2(3M^2-\tilde{M}^2)]}\sigma_1 \nonumber\\
 & &
  -\frac{9H_0(9M^2-2\tilde{M}^2)}{\tilde{M}^4-27\,H_0^2(3\,M^2-\tilde{M}^2)}\Sigma_1
\,,\nonumber\\
\frac{\dot{\Sigma}_1}{N_0} &=& 3\,H_0 \left(\frac{M^2\,\tilde{M}^2\,(9\,H_0^2+\tilde{M}^2)}{H_0\left[\tilde{M}^4+9\,H_0^2(3\,M^2-\tilde{M}^2)\right]}\sigma_1 - \Sigma_1\right)\,.
\end{eqnarray}
Using these, we can express all order ${\cal O}(\epsilon)$ physical quantity in terms of $\sigma_1$ and $\Sigma_1$ only.

Finally, we present the expansion of the $J$ functions used extensively throughout the text. We use the above expansion procedure in Eqs.(\ref{eq:jabdef}) and (\ref{eq:j12def}), to obtain
\begin{eqnarray}
J_\phi^{(x)} &=& 
\frac{2\,H_0^2\,\tilde{M}^2\,e^{3\,\sigma_0}(1-e^{3\,\sigma_0})^2}{2\,(1-H_0)^2\,\tilde{M}^2\,e^{3\,\sigma_0}+9\,M^2(1-H_0\,e^{3\,\sigma_0})^2}+{\cal O}(\epsilon)\,,\nonumber\\
J_\phi^{(y)} &=& \frac{9\,H_0^2\,\tilde{M}^2\,e^{3\,\sigma_0}(1-e^{3\,\sigma_0})
\,\left[3\,M^2\,(9\,H_0^2+\tilde{M}^2)\sigma_1 + 2\,H_0\,(9\,M^2-2\,\tilde{M}^2)\Sigma_1\right]
}{\left[\tilde{M}^4-27\,H_0^2\,(3\,M^2-\tilde{M}^2)\right]\left[2\,(1-H_0)^2\,\tilde{M}^2\,e^{3\,\sigma_0}+9\,M^2(1-H_0\,e^{3\,\sigma_0})^2\right]}+{\cal O}(\epsilon^2)\,,\nonumber\\
J_1 &=& \frac{9\,H_0^2\,M^2(1-e^{3\,\sigma_0})^2}{2\,(1-H_0)^2\,\tilde{M}^2\,e^{3\,\sigma_0}+9\,M^2(1-H_0\,e^{3\,\sigma_0})^2} +{\cal O}(\epsilon)\,,\nonumber\\
J_2 &=& -\frac{27\,H_0^2\,M^2\tilde{M}^2\,e^{3\,\sigma_0}(1-e^{3\,\sigma_0})(9\,H_0^2+\tilde{M}^2)\sigma_1
}{\left[\tilde{M}^4+9\,H_0^2\,(3\,M^2-\tilde{M}^2)\right]\left[2\,(1-H_0)^2\,\tilde{M}^2\,e^{3\,\sigma_0}+9\,M^2(1-H_0\,e^{3\,\sigma_0})^2\right]}+{\cal O}(\epsilon^2)\,.
\end{eqnarray}
\subsection{Odd perturbations}
We now consider the odd sector action in
Eqs.(\ref{eq:oddaction})-(\ref{eq:oddcoeffs}) and expand the background
quantities around the anisotropic fixed point solution, as prescribed in the
previous subsection. The coefficients in the action now read, 
\begin{eqnarray}
K_{11} &=& \frac{a_0^4\,e^{-4\,\sigma_0}\,p_L^2\,p_T^4}{2\,p^2} + {\cal O}(\epsilon)\,,\nonumber\\
K_{22} &=& -\frac{3\,\tilde{M}^2\,e^{2\,\sigma_0}\,a_0^4\,p_T^2\,\left[3\,M^2\,(9\,H_0^2+\tilde{M}^2)\sigma_1 + 2\,H_0\,(9\,M^2-2\,\tilde{M}^2)\Sigma_1\right]}{4\,(1-e^{6\,\sigma_0})\,\left[\tilde{M}^4-27\,H_0^2\,(3\,M^2-\tilde{M}^2)\right]}+{\cal O}(\epsilon^2)\,,
\nonumber\\
\gamma &=& \frac{12\,p_L^2\,p_T^2\,\Sigma_1}{p_T^4} + {\cal O}(\epsilon^2)\,,\nonumber\\
\Omega^2_{11} &=& K_{11}\,\times \left(p^2-\frac{3\,M^2\,p_L^2}{p^2}\right)+ {\cal O}(\epsilon)\,,\nonumber\\
\Omega^2_{12} &=& -\frac{3\,M^2\,a_0^4\,p_L^2\,p_T^4}{4\,e^{4\,\sigma_0}\,p^2}+ {\cal O}(\epsilon)\,,\nonumber\\
\Omega^2_{22} &=& -\frac{3\,M^2\,a_0^4\,p_T^4}{8\,e^{4\,\sigma_0}}
+ {\cal O}(\epsilon)\,.
\label{eq:K22ref}
\end{eqnarray}
where the components of the physical momentum are now defined at order ${\cal O}(\epsilon^0)$, i.e.
\begin{equation}
p_L = \frac{k_L}{a_0\,e^{2\,\sigma_0}}\,,\qquad
p_T = \frac{k_T}{a_0\,e^{-\sigma_0}}\,.
\end{equation}

After the small anisotropy expansion, we see that the first derivative mixing term $\gamma\,K_{22} \sim {\cal O}(\epsilon^2)$ and can be neglected. We introduce a new field basis
\begin{equation}
\bar{\cal V}_1 = \sqrt{K_{11}}\,{\cal V}_1\,,
\qquad
\bar{\cal V}_2 = \sqrt{K_{22}}\,{\cal V}_2\,,
\end{equation}
and bring the kinetic terms to canonical form. Note that we assume that there are no ghost degrees, i.e. $K_{11}, K_{22}>0$. Since the contribution from the time derivatives of the rescaling to the mass term is suppressed in the $\epsilon$ expansion, the odd sector action at leading order becomes
\begin{equation}
I_{\rm odd}^{(2)} \simeq \frac{M_p^2}{2}\,\int N\,dt\,dk_L\,d^2k_T\,a^3
\left[
\frac{\left\vert \dot{\bar{{\cal V}}}_1\right\vert^2}{N^2}
+\frac{\left\vert \dot{\bar{{\cal V}}}_2\right\vert^2}{N^2}
- \frac{\Omega_{11}^2}{K_{11}}\,\left\vert\bar{{\cal V}}_1\right\vert
- \frac{\Omega_{12}^2}{\sqrt{K_{11}}\sqrt{K_{22}}}\,(\bar{\cal V}_1^\star\bar{\cal V}_2+\bar{\cal V}_2^\star\bar{\cal V}_1)
- \frac{\Omega_{22}^2}{K_{22}}\,\left\vert\bar{{\cal V}}_2\right\vert
\right]\,.
\end{equation}
It is then straightforward to find the leading order terms in the dispersion relations,
\begin{eqnarray}
\omega_1^2 &=& p^2 + {\cal O}(\epsilon)\,,\nonumber\\
\omega_2^2 &=& \frac{M^2\,p_T^2(e^{-6\,\sigma_0}-1)\,\left[\tilde{M}^4-27\,H_0^2(3\,M^2-\tilde{M}^2)\right]}{2\,\tilde{M}^2\,\left[3\,M^2\,(9\,H_0^2+\tilde{M}^2)\,\sigma_1 + 2\,H_0\,(9\,M^2-2\,\tilde{M}^2)\,\Sigma_1\right]}+{\cal O}(\epsilon^0)
\,.
\end{eqnarray}
We see that the dispersion relations for small deviation from the fixed point, are dominated by momentum terms. Therefore, although the kinetic term of one of the odd modes tends to vanish on the fixed point, we cannot, in general, integrate out this mode from the action since no mass-gap is present. 

\subsection{Even perturbations}
\label{evenanis}
The kinetic matrix $\bar{K}$ defined in (\ref{eq:defKMO}) is considerably bulky for presentation. In terms of the leading order in small anisotropy expansion around the fixed point, the matrix is formally
\renewcommand{\arraystretch}{1.5}
\begin{equation}
\bar{K} = \left(
\begin{array}{ccc}
{\cal O}(\epsilon^0) & {\cal O}(\epsilon^0) & {\cal O}(\epsilon)\\
{\cal O}(\epsilon^0) & {\cal O}(\epsilon^0) & {\cal O}(\epsilon)\\
{\cal O}(\epsilon) & {\cal O}(\epsilon) & {\cal O}(\epsilon)
\end{array}
\right)\,.
\end{equation}
Since $\bar{K}_{11}\,\bar{K}_{22} - \bar{K}_{12}^2 = {\cal O}(\epsilon^0)$, there is only one mode with a vanishing kinetic term on the fixed point. The two ${\cal O}(\epsilon^0)$ eigenvalues involve square-roots (since they are solutions to a quadratic equation) and are not suitable for presentation (nor calculation). However, going to a new basis with a non-orthogonal transformation, we can have (relatively) simpler eigenvalues. We introduce a transformation of the form:
\begin{equation}
R = \left(
\begin{array}{ccc}
1  & 0 & 0 \\
\frac{\bar{K}_{12}\,\bar{K}_{33} - \bar{K}_{13}\,\bar{K}_{23}}{\bar{K}_{23}^2 -\bar{K}_{22}\,\bar{K}_{33}} & 1 & 0\\
\frac{\bar{K}_{13}\,\bar{K}_{22} - \bar{K}_{12}\,\bar{K}_{23}}{\bar{K}_{23}^2 -\bar{K}_{22}\,\bar{K}_{33}} & -\frac{\bar{K}_{23}}{\bar{K}_{33}} & 1
\end{array}
\right)\,,
\end{equation}
which we use to define new basis ${\cal Y} = R\,\bar{\cal Y}$. In this basis, the kinetic matrix becomes:
\begin{equation}
R^T\,\bar{K}\,R = 
\left(
\begin{array}{ccc}
\kappa_1 & 0 & 0\\
0 & \kappa_2 & 0\\
0 & 0 & \kappa_3
\end{array}
\right) = 
\left(
\begin{array}{ccc}
\bar{K}_{11} + 
\frac{\bar{K}_{13}^2\,\bar{K}_{22}+\bar{K}_{12}^2\,\bar{K}_{33}-2\,\bar{K}_{12}\,\bar{K}_{13}\,\bar{K}_{23}}{\bar{K}_{23}^2-\bar{K}_{22}\,\bar{K}_{33}}
& 0 & 0 \\
0 & \bar{K}_{22} - \frac{\bar{K}_{23}^2}{\bar{K}_{33}} &0\\
0 & 0 & \bar{K}_{33}
\end{array}
\right)\,.
\end{equation}
Notice that the eigenvalues $\kappa_i$ are \emph{not} the same as the eigenvalues of the matrix $\bar{K}$, since the transformation $R$ is not orthogonal. However, $R$ has unit determinant, so the determinant of the kinetic matrix does not change. This is enough to study the signature of the eigenvalues. The eigenvalues can then be computed to be:
\begin{eqnarray}
\kappa_1 &=& \left[ \frac{8\,p^4}{p_T^4} - \frac{8\,\tilde{M}^4}{\tilde{M}^4+9\,H_0^2(3\,M^2-\tilde{M}^2)}\right]^{-1}
+{\cal O}(\epsilon)\,,\nonumber\\
\kappa_2 &=& \frac{2\,a_0^4\,e^{8\,\sigma_0}\,\tilde{M}^2\,p_L^2\,\left[9\,H_0^2\,p^4\,(\tilde{M}^2-3\,M^2)+\tilde{M}^4\,p_L^2\,(-2\,p^2+p_L^2)\right]}{\tilde{M}^2\,p_L^2(\tilde{M}^2-3\,p^2)^2-9 H_0^2(\tilde{M}^2-3\,M^2)\,\left[6\,p^4+\tilde{M}^2(-4\,p^2+p_L^2)\right]}
+{\cal O}(\epsilon)\,,\nonumber\\
\label{eq:kinvan}\kappa_3 &=& -\frac{3\,\tilde{M}^2\,e^{2\,\sigma_0}\,a_0^4\,p_T^2\,\left[3\,M^2\,(9\,H_0^2+\tilde{M}^2)\sigma_1 + 2\,H_0\,(9\,M^2-2\,\tilde{M}^2)\Sigma_1\right]}{(1-e^{6\,\sigma_0})\,\left[\tilde{M}^4-27\,H_0^2\,(3\,M^2-\tilde{M}^2)\right]}+{\cal O}(\epsilon^2)\,,\,.
\label{kappas}
\end{eqnarray}
The eigenvalues which are non-zero on the fixed point yield the following no-ghost condition on the parameters
\begin{equation}
\kappa_1 > 0\,,
\qquad
\kappa_2 >0\,.
\end{equation}
The third eigenvalue, which is of order ${\cal O}(\epsilon)$, is actually proportional to the kinetic term of a similar mode in the odd sector, with
\begin{equation}
\kappa_3 = 4\,K_{22}^{\rm odd}+{\cal O}(\epsilon)^2\,,
\end{equation}
where $K_{22}^{\rm odd}$ is given in the second line of Eq.(\ref{eq:K22ref}). Satisfying the third no ghost condition $\kappa_3>0$ is rather difficult, since it involves evolving quantities $\sigma_1$ and $\Sigma_1$.

\subsection{Evading the ghosts}

Since we are looking for a region in the parameter space where none of the degrees of freedom are ghosts, we impose that the eigenvalues of the kinetic matrix in the even sector are positive, i.e.
\begin{equation}
\kappa_1,\, \kappa_2,\, \kappa_3 >0\,,
\end{equation}
which ensures that the odd sector is also safe.

For the first two kinetic terms in Eq.(\ref{kappas}), it is possible to find regions where a given mode with any momentum is not a ghost. The sufficient (but not necessary) conditions we adopt here are,
\begin{equation}
\tilde{M}^2 < 0 \,, \qquad
M^2 < \frac{\tilde{M}^2(9 H_0^2-\tilde{M}^2)}{27\,H_0^2} < 0\,.
\label{noghost1}
\end{equation}
Thus, $\kappa_1$ and $\kappa_2$ can be made positive, regardless of the momentum of the modes.

With a parameter set satisfying (\ref{noghost1}), $\kappa_3>0$ gives
\begin{equation}
{\rm Sgn}(1-e^{\sigma_0})\,\left[-3\,\vert M^2\vert\,\left(9\,H_0^2-\vert \tilde{M}^2\vert\right)\sigma_1 +2\,H_0\,\left(2\,\vert \tilde{M}^2\vert-9\,\vert M^2\vert\right)\Sigma_1\right]>0\,,
\label{noghost3}
\end{equation}
which depends linearly on the deviation of the background from its fixed point, i.e. the two dynamical  functions $\sigma_1$ and $\Sigma_1$. This means that regardless of the value of $M^2$ and $\tilde{M}^2$, there may always be a region where $\sigma_1$ and $\Sigma_1$ conspire to give a negative $\kappa_3$. On the other hand, close to the attractor, the evolution of these two quantities depend on each other. The equation of motion for $\sigma_1$ can be written as \cite{Gumrukcuoglu:2012aa}
\begin{equation}
\dot{\Sigma}_1 +3\,H_0 \,\Sigma_1 + M_\sigma^2\,\sigma_1 = 0\,,
\label{eq:sigma1evol}
\end{equation}
where
\begin{equation}
M_\sigma ^2 \equiv-\frac{3\,M^2\,\tilde{M}^2\,(9\,H_0^2+\tilde{M}^2)}{\tilde{M}^4+9\,H_0^2(3\,M^2-\tilde{M}^2)}
\,.
\end{equation}
First, the condition that the fixed point is stable, i.e. $M_\sigma^2>0$, combined with the conditions (\ref{noghost1}), yields
\begin{equation}
9 H_0^2 - \vert\tilde{M}^2\vert >0\,.
\label{fixedpointstability}
\end{equation}
We also wish that $\sigma_1$ is over-damped, so that there are no out-of-control changes of signs:
\begin{equation}
9\,H_0^2> 4\,M_\sigma^2\,.
\label{noghost2}
\end{equation}
The solutions to Eq.(\ref{eq:sigma1evol}) are easy to obtain,
\begin{equation}
\sigma = e^{-\frac{3}{2}\,H_0\,t}\left[C_1 \,e^{\sqrt{\frac{9}{4}H_0^2 -M_\sigma^2}t} + C_2 \,e^{-\sqrt{\frac{9}{4}H_0^2 -M_\sigma^2}t}\right]\,.
\end{equation}
At late times, the general solution approaches to the $C_2=0$ solution, where we can write
\begin{equation}
\left.\Sigma_1 \right\vert_{t\to\infty} = \left(-\frac{3}{2}H_0 + \sqrt{\frac{9}{4}\,H_0^2-M_\sigma^2}\right)\left.\sigma_1 \right\vert_{t\to\infty} \,.
\label{lateattractor}
\end{equation}
So, in this regime, the condition (\ref{noghost3}) can in principle be satisfied by choosing the appropriate sign for $\sigma_1$.

As an example, we choose the set of parameters used in \cite{Gumrukcuoglu:2012aa}
\begin{equation}
\alpha_3 = -\frac1{20} \,,\qquad
\alpha_4 = 1\,,\qquad
\Lambda =0\,,\qquad
\mu = 20\,,
\label{paramset}
\end{equation}
implying
\begin{equation}
e^\sigma \simeq 0.51 \,,\qquad
H_0 \simeq 15.35 \,.
\end{equation}
As a result, we find,
\begin{equation}
M^2 = -45079\,,\quad
\tilde{M}^2 = -460.66\,,\quad
\frac{\tilde{M}^2\,(9\,H_0^2-\tilde{M}^2)}{27\,H_0^2} = -186.9\,,\quad
M_\sigma^2 = 362.10\,,\quad
\frac{9}{4}H_0^2- M_\sigma^2 = 168.10\,,
\end{equation}
which satisfy the no-ghost conditions for the first two modes (\ref{noghost1}), fixed point stability condition (\ref{fixedpointstability}), and over-damping condition (\ref{noghost2}). For this example, the kinetic term for the third mode, along with the late time attractor assumption (\ref{lateattractor}) becomes
\begin{equation}
\kappa_3 \simeq -41.85\,a_0^4\,p_T^2\,\sigma_1\,.
\end{equation}
Thus, the parameter set (\ref{paramset}) does not lead to any ghost degree, provided that \emph{i.} we are close to the late time attractor; \emph{ii.} we are in the correct side of the fixed point, i.e. $\sigma_1<0$ .

\section{Discussion}
\label{sec:discussion}

Since the introduction of the dRGT theory, there has been several
attempts to construct a cosmological solution that is fully consistent
with the known expansion history of the universe. In the present paper,
we have shown that the self-accelerating solutions introduced in
\cite{Gumrukcuoglu:2011ew, Gumrukcuoglu:2011zh} have a ghost instability
at nonlinear order. A technical source of this problem is that the
kinetic terms of three among the five degrees of freedom are exactly
proportional to the equation of motion of the temporal St\"uckelberg
field for the self-accelerating branch. These kinetic terms reappear in
the cubic order, and their behavior can be understood by a deformation 
of the FLRW symmetries. These results were previously reported in
\cite{DeFelice:2012mx} and are also compatible with the independent
analysis of Ref. \cite{D'Amico:2012pi}.

The next step is to seek another class of cosmological solutions on
which linear perturbations have non-zero (and positive) kinetic
terms. To this end, one needs to ensure that the factorized form of the
St\"uckelberg equation of motion is broken, since kinetic terms of some
perturbation degrees of freedom are typically proportional to one of the
factors in this equation. Since the FLRW symmetry consists of
homogeneity and isotropy, there are two possibilities: breaking either
homogeneity or isotropy.

In the present paper we have considered the possibility of breaking the
FLRW symmetry by introducing anisotropy. With a relatively large
anisotropy, we found a healthy region with non-zero and positive kinetic
terms for all five perturbation degrees of freedom. This healthy region
is in a neighborhood of (but not exactly on) the de Sitter fixed point
solution with an anisotropic hidden sector introduced previously in 
Ref.~\cite{Gumrukcuoglu:2012aa}. To be more precise, signatures of the 
resulting kinetic terms depend not only on the direction of deformation
of the background from the fixed point, but also on the signature of its
time derivative. If the initial condition is such that the evolution
is close to the fixed point and that the deformation is in the correct
direction, then all five graviton polarizations in the anisotropic 
background can have positive kinetic terms for a range of parameters.

Exactly on the fixed point, however, quadratic kinetic terms for two of
the five degrees vanish, while all other three degrees have positive
kinetic terms in a finite region in the parameter space. Hence, for a
small deviation from the fixed point, even though appearance of the
ghost degrees can be avoided, two degrees of freedom have very small
(cubic order) kinetic terms and no mass gap (see Appendix \ref{app:dispIR}). 
This is an indication that these degrees are strongly coupled 
and remains to be a source of concern.

Nevertheless, the anisotropic FLRW solution studied here is the first
calculable example of a stable cosmology in the dRGT theory of nonlinear
massive gravity. One of technical advantages of this solution is that
the spatial homogeneity and the SO(2) invariance of the axisymmetric 
background allows decoupling between even and odd sectors at the linear
order. Moreover, the absence of ghost in a neighborhood of this solution
may indicate the existence of region without ghost nor strong coupling
somewhere between this fixed point solution and the isotropic FLRW
solution. 
We also note that the perturbations around other Bianchi type universes remain unexplored.

Alternatively, one may decide to break the FLRW symmetries by introducing
inhomogeneities. For instance, there are such known solutions 
\cite{D'Amico:2011jj, Koyama:2011xz, Koyama:2011yg, Gratia:2012wt, Kobayashi:2012fz,
Volkov:2012cf, Motohashi:2012jd}, where the background dynamics
is FLRW--like (or reduces to FLRW through cosmological Vainshtein
mechanism \cite{D'Amico:2011jj}) while the St\"uckelberg sector is contaminated by spatial 
inhomogeneities. In principle, these may also evade the FLRW ghost, 
although in the lack of symmetry, the stability analysis of
perturbations becomes technically challenging (although see
\cite{Wyman:2012iw}). 

A different approach to break the factorized form of the time
St\"uckelberg equation is to extend the theory, through introduction of
new dynamics, e.g. by imposing a dilatation-like symmetry
\cite{D'Amico:2012zv} or considering variations of the parameters with a
scalar field \cite{Huang:2012pe}. Indeed, the resulting cosmological 
solutions \cite{Wu:2013ii, Hinterbichler:2013dv, Leon:2013qh} can still 
be fully homogeneous and isotropic, with non-zero kinetic terms for 
perturbations. The perturbation analysis of these
extensions will be addressed in a separate work \cite{Gumrukcuoglu:2013nza}.

\begin{acknowledgments}
We would like to thank Gianmassimo Tasinato for useful discussions. The work of A.E.G, C.L. and S.M. was supported by the World Premier International Research Center Initiative (WPI Initiative), MEXT,
 Japan. S.M. also acknowledges the support by Grant-in-Aid for
 Scientific Research 17740134, 19GS0219, 21111006, 21540278, by 
 Japan-Russia Research Cooperative Program.  
\end{acknowledgments}

\appendix

\section{Nonexistence of less symmetric fixed point solution of Bianchi--I}

In this Appendix, we show that fixed point solutions of Bianchi--I type
always have axisymmetry, as long as the expansion is constrained to be isotropic and the fiducial metric is de Sitter. 
We introduce the generic Bianchi--I metric as
\begin{equation}
ds^2 = - N^2\,dt^2 + \bar{a}^2 \left[e^{2(\sigma_1+\sigma_2)}\,dx^2 +e^{-2\,\sigma_1}\,dy^2 + e^{-2\,\sigma_2}\,dz^2\right]\,,
\label{metric}
\end{equation}
which, for $\sigma_1 = \sigma_2$, $\sigma_1=-2\,\sigma_2$ and $\sigma_2=-2\,\sigma_1$, reduces to the axisymmetric metric used in \cite{Gumrukcuoglu:2012aa}, up to redefinition of coordinates. 

For the present discussion, the equations of motion for $\sigma_1$ and $\sigma_2$ are sufficient:

\begin{eqnarray}
\frac{1}{3\,N}\left(2\,\dot{\Sigma}_1+\dot{\Sigma}_2\right)+H\left(2\,\Sigma_1+\Sigma_2\right) &=&
\frac{m_g^2}{3}\,\left(e^{-\sigma_1-\sigma_2}-e^{\sigma_1}\right)\,X\,
\left[ (3+3\alpha_3+\alpha_4)-(1+2\alpha_3+\alpha_4)\,(e^{\sigma_2}  +r)\,X
\right.\nonumber\\
&&\left.\qquad\qquad\qquad\qquad\qquad\qquad\qquad\qquad\qquad\qquad
+ (\alpha_3+\alpha_4)\,e^{\sigma_2}\,r\,X^2\right]\,,\nonumber\\
\frac{1}{3\,N}\left(\dot{\Sigma}_1+2\,\dot{\Sigma}_2\right)+H\left(\Sigma_1+2\,\Sigma_2\right) &=&
\frac{m_g^2}{3}\,\left(e^{-\sigma_1-\sigma_2}-e^{\sigma_2}\right)\,X\,
\left[ (3+3\alpha_3+\alpha_4)-(1+2\alpha_3+\alpha_4)\,(e^{\sigma_1}  +r)\,X
\right.\nonumber\\
&&\left.\qquad\qquad\qquad\qquad\qquad\qquad\qquad\qquad\qquad\qquad
+ (\alpha_3+\alpha_4)\,e^{\sigma_1}\,r\,X^2\right]\,,\label{generalbianchi}
\end{eqnarray}
where
\begin{equation}
r \equiv \frac{n}{N\,X}\,,\qquad
X \equiv \frac{\alpha}{a}\,,\qquad
H \equiv \frac{\dot{a}}{a\,N}\,,\qquad
\Sigma_1 \equiv \frac{\dot{\sigma}_1}{N}\,,\qquad
\Sigma_2 \equiv \frac{\dot{\sigma}_2}{N}\,.
\end{equation}
We now look for fixed points of the type found in \cite{Gumrukcuoglu:2012aa}, namely, which have isotropy in the expansion ($\Sigma_1=\Sigma_2=0$), but not in the normalization ($\sigma_1\neq0,\,\sigma_2 \neq 0$). Under these conditions, Eq.~(\ref{generalbianchi}) can be written as
\begin{eqnarray}
\left(e^{-\sigma_1-\sigma_2}-e^{\sigma_1}\right)\,
\left[ (3+3\alpha_3+\alpha_4)-(1+2\alpha_3+\alpha_4)\,(e^{\sigma_2}  +r)\,X+ (\alpha_3+\alpha_4)\,e^{\sigma_2}\,r\,X^2\right]&=&0\,,\label{eqs1b}\\
\left(e^{-\sigma_1-\sigma_2}-e^{\sigma_2}\right)\,
\left[ (3+3\alpha_3+\alpha_4)-(1+2\alpha_3+\alpha_4)\,(e^{\sigma_1}  +r)\,X+ (\alpha_3+\alpha_4)\,e^{\sigma_1}\,r\,X^2\right]&=&0\,.\label{eqs2b}
\end{eqnarray}
Let us discuss the solutions to these equations. For Eq.(\ref{eqs1b}), one of the solutions is
\begin{equation}
e^{\sigma_2}=e^{-2\,\sigma_1}\,,
\end{equation}
implying axial symmetry around $\hat{z}$ direction. Similarly, Eq.(\ref{eqs2b}) has a solution
\begin{equation}
e^{\sigma_1}=e^{-2\,\sigma_2}\,,
\end{equation}
which implies axial symmetry around $\hat{y}$ direction. Both of these solutions correspond to the solutions already studied in \cite{Gumrukcuoglu:2012aa}. Removing these, (\ref{eqs1b}) and (\ref{eqs2b}) become
\begin{eqnarray}
(3+3\alpha_3+\alpha_4)-(1+2\alpha_3+\alpha_4)\,(e^{\sigma_2}  +r)\,X+ (\alpha_3+\alpha_4)\,e^{\sigma_2}\,r\,X^2 &=& 0\,, \nonumber\\
(3+3\alpha_3+\alpha_4)-(1+2\alpha_3+\alpha_4)\,(e^{\sigma_1}  +r)\,X+ (\alpha_3+\alpha_4)\,e^{\sigma_1}\,r\,X^2&=&0\,,
\end{eqnarray}
for which, the solutions $\sigma_1$ and $\sigma_2$ are the same, implying axial symmetry around $\hat{x}$ direction. The only alternative solution is $X = (2+\alpha_3)/[r\,(1+\alpha_3)]$, which is realized only for $\alpha_4=1+\alpha_3+\alpha_3^2$. Moreover, the remaining equations require a further tuning between $\alpha_3$, $m_g$ and $\Lambda$, so we drop this solution as well.

Thus, we conclude that all the de Sitter fixed points in Bianchi--I with isotropic expansion are axisymmetric.

\section{Explicit action for even modes around axisymmetric Bianchi--I}
\label{sec:evenact}
In this Appendix, we present the explicit expression for the action quadratic in even modes. After switching to momentum space, the action has the form (\ref{eq:acteven1})
\begin{equation}
I^{(2)}_{\rm even} = \frac{M_p^2}{2}\,\int N\,dt\,dk_L\,d^2k_T\,a^3 \left[\frac{\dot{{\cal Y}}^\dagger}{N} \,K\, \frac{\dot{{\cal Y}}}{N} -{\cal Y}^\dagger\, \Omega^2\,{\cal Y} + {\cal Z}^\dagger \,{\cal A}\,{\cal Y} + {\cal Y}^\dagger \,{\cal A}^T\,{\cal Z} +
{\cal Z}^\dagger \,{\cal B}\,\frac{\dot{\cal Y}}{N}+ \frac{\dot{\cal Y}^\dagger}{N}\,{\cal B}^T\,{\cal Z} + {\cal Z}^\dagger\,{\cal C}\,{\cal Z}
\right]\,,
\end{equation}
where the two field vectors are 
\begin{equation}
{\cal Y}  = \left(
\begin{array}{ccc}
 \hat{\psi}\\
 \hat{\beta}\\
 \hat{E}_\pi
\end{array}
\right)\,,
\qquad
{\cal Z} = \left(
\begin{array}{ccc}
 \hat{\Phi}\\
 \hat{B}\\
 \hat{\chi}\\
 \hat{\tau}_\pi
\end{array}
\right)\,,
\end{equation}
while $K$ and $\Omega^2$ are $3\times3$ real symmetric matrices, ${\cal C}$ is a $4\times 4$ real symmetric matrix, ${\cal A}$ and ${\cal B}$ are $4\times 3$ real matrices. Up to background equations, the components of these matrices are
\renewcommand{\arraystretch}{1.5}
\begin{equation}
K = m_g^2\,a^4\,X\,
\left(
\begin{array}{ccc}
0 & 0 & 0 \\
0 & \frac{e^{6\,\sigma}\,J_\phi^{(x)}\,p_L^2}{1+e^{2\,\sigma}\,r}  & 0\\
0 & 0 & \frac{e^{-2\,\sigma}\,J_\phi^{(y)}\,p_T^2}{e^{\sigma}+r}
\end{array}
\right)
\end{equation}
\begin{equation}
\Omega^2 =
\frac{m_g^2\,a^4\,e^{2\,\sigma}\,X\,p_T^2}{e^{6\,\sigma}-1}
\left[
(e^{2\,\sigma}\,r-1)\,J_\phi^{(x)}-e^{2\,\sigma}\,(r-e^\sigma)\,J_\phi^{(y)} \right]
\left(
\begin{array}{ccc}
0 & 0& -\frac{1+e^{-3\,\sigma}}{2\,a^2}\\
0 & e^{4\,\sigma}\,p_L^2 & -e^\sigma\,p_L^2\\
-\frac{1+e^{-3\,\sigma}}{2\,a^2} & -e^\sigma\,p_L^2 & e^{-2\,\sigma}\,p_L^2
\end{array}
\right)
\end{equation}
\begin{equation}
{\cal A} = \left(
\begin{array}{ccc}
{\cal A}_{11} & {\cal A}_{12} & {\cal A}_{13} \\
{\cal A}_{21} & 0 & 0 \\
0 & 0 & 0 \\
{\cal A}_{41} & {\cal A}_{42} & {\cal A}_{43}
\end{array}
\right)\,,
\qquad
{\cal B} = \left(
\begin{array}{ccc}
{\cal B}_{11} & 0 & 0\\
{\cal B}_{21} & 0 & {\cal B}_{23} \\
0 & {\cal B}_{32} & 0 \\
{\cal B}_{41} & {\cal B}_{42} & {\cal B}_{43}
\end{array}
\right)\,,
\qquad
{\cal C} = \left(
\begin{array}{cccc}
{\cal C}_{11} & {\cal C}_{12} & {\cal C}_{13}& {\cal C}_{14}\\
{\cal C}_{12} & {\cal C}_{22} & {\cal C}_{23}& {\cal C}_{24}\\
{\cal C}_{13} & {\cal C}_{23} & {\cal C}_{33}& {\cal C}_{34}\\
{\cal C}_{14} & {\cal C}_{24} & {\cal C}_{34}& {\cal C}_{44}
\end{array}
\right)\,,
\end{equation}
where
\begin{eqnarray}
{\cal A}_{11} &=& \frac{1}{2}\left(p_T^2 +m_g^2\,e^{-2\,\sigma}\,X\,J_\phi^{(x)}\right)\,, \nonumber\\
{\cal A}_{12} &=&  m_g^2 \,a^2\,e^{2\,\sigma}\,X\,J_\phi^{(x)}\,p_L^2\,, \nonumber\\
{\cal A}_{13} &=&  m_g^2 \,a^2\,e^{-\sigma}\,X\,J_\phi^{(y)}\,p_T^2\,, \nonumber\\
{\cal A}_{21} &=& -\frac{3}{2}\,a\,e^{-\sigma}\,\Sigma\,p_T^2\,,\nonumber\\
{\cal A}_{41} &=& \frac{3\,m_g^2\,N\,r\,X\,J_\phi^{(x)}}{2\,(1-e^{3\,\sigma})}\,\left(H-H_f\,e^\sigma\,X\right)\,,\nonumber\\
{\cal A}_{42} &=& \frac{3\,m_g^2\,a^2\,e^{4\,\sigma}\,N\,r\,X\,p_L^2\,J_\phi^{(x)}}{1-e^{3\,\sigma}}\,\left(H-H_f\,e^\sigma\,X\right)\,,\nonumber\\
{\cal A}_{43} &=& \frac{m_g^2\,a^2\,e^{-\sigma}\,N\,r\,X\,p_T^2}{e^\sigma-r}
\left[J_\phi^{(y)}\,\left(H-\Sigma -H_f\,r\,X\right) +J_1\,\left(H-\Sigma -H_f\,e^\sigma\,X\right) + J_2\,\left(H+2\,\Sigma-H_f\,e^{-2\,\sigma}\,X\right)\right]\,,
\end{eqnarray}
\begin{eqnarray}
{\cal B}_{11} &=& H-\Sigma\,,\nonumber\\
{\cal B}_{21} &=& -\frac{1}{2}\,a\,e^{-\sigma}\,p_T^2\,,\nonumber\\
{\cal B}_{23} &=& -\frac{m_g^2\,a^3\,e^{-\sigma}\,X\,J_\phi^{(y)}\,p_T^2}{e^\sigma+r}\,,\nonumber\\
{\cal B}_{32} &=& -\frac{m_g^2\,a^3\,e^{4\,\sigma}\,X\,J_\phi^{(x)}\,p_L^2}{1+e^{2\,\sigma}\,r}\,,\nonumber\\
{\cal B}_{41} &=& \frac{1}{2}\,m_g^2\,N\,r\,X\,J_\phi^{(x)}\,,\nonumber\\
{\cal B}_{42} &=&  \frac{m_g^2\,a^2\,e^{4\,\sigma}\,N\,r\,X\,J_\phi^{(x)}\,p_L^2}{1+e^{2\,\sigma}\,r}\,,\nonumber\\
{\cal B}_{43} &=&  \frac{m_g^2\,a^2\,e^{-\sigma}\,N\,r\,X\,J_\phi^{(y)}\,p_T^2}{e^{\sigma}+r}\,,
\end{eqnarray}
\begin{eqnarray}
{\cal C}_{11} &=& -6\,(H^2- \Sigma^2)\,,\nonumber\\
{\cal C}_{12} &=& a\,e^{-\sigma}\,p_T^2\,(2\,H+\Sigma)\,,\nonumber\\
{\cal C}_{13} &=& 2\,a\,e^{2\,\sigma}\,p_L^2\,(H-\Sigma)\,,\nonumber\\
{\cal C}_{14} &=&- m_g^2\,e^{-2\,\sigma}\,N\,r\,X^2\,H_f\,\left(J_\phi^{(x)}+2\,e^{3\,\sigma}\,J_\phi^{(y)}\right)\,,\nonumber\\
{\cal C}_{22} &=& \frac{1}{2}\,a^2\,p_T^2\,\left( e^{-2\,\sigma}\,p_L^2 + \frac{2\,m_g^2\,X\,J_\phi^{(y)}}{e^\sigma+r}\right)\,,\nonumber\\
{\cal C}_{23} &=& -\frac{1}{2}\,a^2\,e^\sigma\,p_L^2\,p_T^2\,,\nonumber\\
{\cal C}_{24} &=& \frac{m_g^2\,a\,e^{-\sigma}\,N\,r^2\,X\,J_\phi^{(y)}\,p_T^2}{e^\sigma+r}\,,\nonumber\\
{\cal C}_{33} &=& \frac{1}{2}\,a^2\,p_L^2\left(e^{4\,\sigma}\,p_T^2+\frac{2\,m_g^2\,X\,J_\phi^{(x)}}{e^{-2\,\sigma}+r}\right)\,,\nonumber\\
{\cal C}_{34} &=& \frac{m_g^2\,a\,e^{4\,\sigma}\,N\,r^2\,X\,J_\phi^{(x)}\,p_L^2}{1+e^{2\,\sigma}\,r}\,,\nonumber\\
{\cal C}_{44} &=& m_g^2\,N^2\,r\,X\,\Bigg\{ r\,\left(\frac{J_\phi^{(x)}\,p_L^2}{e^{-2\,\sigma}+r}+\frac{J_\phi^{(y)}\,p_T^2}{e^{\sigma}+r}\right) -2\,(H-\Sigma)^2\,\left(\frac{e^\sigma}{e^\sigma-r}\,J_1  + J_\phi^{(y)} \right)\nonumber\\
&&\qquad 
-\frac{H-\Sigma}{e^\sigma-r}\,\left[(H+2\,\Sigma)\left(2\,e^\sigma\,J_2+2\,e^{-2\,\sigma}\,(J_2-J_\phi^{(x)})-J_\phi^{(x)}\,r\right)
-H_f\,e^{-2\,\sigma}\,X\,\left(2\,e^{4\,\sigma}\,J_1+2\,e^\sigma\,J_2-J_\phi^{(x)}\,r\right)
\right]\nonumber\\
&&\qquad
-H_f\,e^{-2\,\sigma}\,X\,(H+2\,\Sigma)\,\left(J_\phi^{(x)}- \frac{2\,e^\sigma\,(J_2-J_\phi^{(x)})}{e^\sigma-r}\right)
-\left(J_\phi^{(x)}+2\,J_\phi^{(y)}\right)\,\frac{\dot{H}}{N}-2\,\left(J_\phi^{(x)}-J_\phi^{(y)}\right)\,\frac{\dot{\Sigma}}{N}
\Bigg\}
\end{eqnarray}

\section{Diagonalizing the action for even modes near FLRW solutions}
\label{app:diagonalize}

In this Appendix, we analyze the quadratic action for even modes around
a Bianchi--I background solution, employing the small anisotropy
expansion. We obtain a simple action for the even modes, through a
series of field redefinitions and transformations.

\subsection{Applying small anisotropy expansion}
We start with the action (\ref{eq:acteven2}) in the small anisotropy expansion as discussed in Sec.\ref{sec:evenexpand},

\begin{equation}
I^{(2)}_{\rm even} = \frac{M_p^2}{2}\,\int N\,dt\,dk_L\,d^2k_T\,a^3 \left[\frac{\dot{{\cal Y}_1}^\dagger}{N} \,K_1\, \frac{\dot{{\cal Y}}_1}{N} +\frac{\dot{{\cal Y}}_1^\dagger}{N}\,M_1\,{\cal Y}_1+{\cal Y}_1^\dagger\,M_1^T\,\frac{\dot{{\cal Y}}_1}{N}
-{\cal Y}_1^\dagger\, \Omega^2_1\,{\cal Y}_1
\right]\,.
\end{equation}
Since we will be applying several transformations, it is useful to keep track of each step, so we change the notation such that ${\cal Y} \to {\cal Y}_1$, $\bar{K}\to K_1$, $\bar{M}\to M_1$ and $\bar{\Omega}^2\to \Omega_1^2$.
The leading order terms of the matrix components are given by
\begin{eqnarray}
\left(K_1\right)_{11} &=& \frac{p_T^4}{8\,p^4} + {\cal O}(\epsilon)\,,\nonumber\\
\left(K_1\right)_{12} &=& -\frac{a^2\,M_{GW}^2\,p_L^2\,(2\,p^2+p_T^2)}{2\,p^4\,(1-r^2)}\,\sigma + {\cal O}(\epsilon^2)\,,\nonumber\\
\left(K_1\right)_{13} &=& -\frac{a^2\,M_{GW}^2\,p_T^2\,(p^2+p_L^2)}{4\,p^4\,(1-r^2)}\,\sigma + {\cal O}(\epsilon^2)\,,\nonumber\\
\left(K_1\right)_{22} &=& - \frac{2\,a^4\,M_{GW}^2\,p_L^2}{1-r^2}\,\sigma + {\cal O}(\epsilon^2)\,,\nonumber\\
\left(K_1\right)_{23} &=& \frac{a^4\,M_{GW}^2\,p_L^2 \,p_T^2\,\left(p^4\,r^2-3\,H^2\,M_{GW}^2\right)}{3\,H^2\,p^4\,(1-r^2)^2}\,\sigma^2+{\cal O}(\epsilon^3)\,,\nonumber\\
\left(K_1\right)_{33} &=& \frac{a^4\,M_{GW}^2\,p_T^2}{1-r^2}\,\sigma +{\cal O}(\epsilon^2)\,.
\end{eqnarray}
\begin{eqnarray}
\left(M_1\right)_{11} &=& \frac{p_T^4}{8\,H\,p^2} +{\cal O}(\epsilon)\,,\nonumber\\
\left(M_1\right)_{12} &=& -\frac{a^2\,M_{GW}^2\,p_L^2}{12\,H\,p^4\,(1-r^2)}\,\left[6\,p^2\,p_T^2 + 2\,p^2\,(2\,p^2+3\,p_T^2)\,r+3\,p_T^4\,r^2\right]\,\sigma + {\cal O}(\epsilon^2)\,,\nonumber\\
\left(M_1\right)_{13} &=& -\frac{a^2\,M_{GW}^2\,p_T^2}{12\,H\,p^4\,(1-r^2)}\,\left[-3\,p^2\,p_T^2 + p^2\,(p^2+3\,p_L^2)\,r+3\,p_T^4\,r^2\right]\,\sigma + {\cal O}(\epsilon^2)\,,\nonumber\\
\left(M_1\right)_{21} &=& -\frac{a^2\,M_{GW}^2\,p_L^2}{6\,H\,p^2\,(1-r^2)}\,\left(2\,p^2\,r + 3\,p_T^2\right)\,\sigma + {\cal O}(\epsilon^2)\,,\nonumber\\
\left(M_1\right)_{22} &=& -\frac{2\,a^4\,M_{GW}^2\,p_L^4\,r}{3\,H\,(1-r^2)}\,\sigma + {\cal O}(\epsilon^2)\,,\nonumber\\
\left(M_1\right)_{23} &=& -\frac{2\,a^4\,M_{GW}^2\,p_L^2\,p_T^2\,r}{3\,H\,(1-r^2)}\,\sigma + {\cal O}(\epsilon^2)\,,\nonumber\\
\left(M_1\right)_{31} &=& \frac{a^2\,M_{GW}^2\,p_T^2}{12\,H\,p^2\,(1-r^2)}\,\left(2\,p^2\,r+3\,p_T^2\right)\,\sigma + {\cal O}(\epsilon^2)\,,\nonumber\\
\left(M_1\right)_{32} &=& \frac{a^4\,M_{GW}^2\,p_L^2\,p_T^2\,r}{3\,H\,(1-r^2)}\,\sigma + {\cal O}(\epsilon^2)\,,\nonumber\\
\left(M_1\right)_{33} &=& \frac{a^4\,M_{GW}^2\,p_T^4\,r}{3\,H\,(1-r^2)}\,\sigma + {\cal O}(\epsilon^2)\,,
\end{eqnarray}
\begin{eqnarray}
\left(\Omega_1^2\right)_{11} &=& \frac{M_{GW}^2}{6}+{\cal O}(\epsilon)\,,\nonumber\\
\left(\Omega_1^2\right)_{12} &=& \frac{1}{3}\,a^2\,M_{GW}^2\,p_L^2+{\cal O}(\epsilon)\,,\nonumber\\
\left(\Omega_1^2\right)_{13} &=& -\frac{1}{6}\,a^2\,M_{GW}^2\,p_T^2+{\cal O}(\epsilon)\,,\nonumber\\
\left(\Omega_1^2\right)_{22} &=& \frac{1}{6}\,a^4\,M_{GW}^2\,p_L^2\,\left(3\,p^2+p_L^2\right)+{\cal O}(\epsilon)\,,\nonumber\\
\left(\Omega_1^2\right)_{23} &=& \frac{1}{6}\,a^4\,M_{GW}^2\,p_L^2\,p_T^2+{\cal O}(\epsilon)\,,\nonumber\\
\left(\Omega_1^2\right)_{33} &=& \frac{1}{6}\,a^4\,M_{GW}^2\,p_T^2\,\left(3\,p^2+p_T^2\right)+{\cal O}(\epsilon)\,,
\end{eqnarray}

Although the mixing matrix $M_1$ can be made anti-symmetric by adding boundary terms, we keep it for now, since the following transformations will spoil its symmetries anyway.

\subsection{Diagonalizing the kinetic matrix}
We start by diagonalizing the kinetic matrix. To achieve this, we apply the following field transformation,
\begin{equation}
{\cal Y}_2 = R_1^{-1}\,{\cal Y}_1\,,
\end{equation}
where the transformation matrix is
\begin{equation}
R_1 = \left(
\begin{array}{ccc}
1  & 0 & 0 \\
\frac{(K_1)_{12}\,(K_1)_{33} - (K_1)_{13}\,(K_1)_{23}}{(K_1)_{23}^2 -(K_1)_{22}\,(K_1)_{33}} & 1 & 0\\
\frac{(K_1)_{13}\,(K_1)_{22} - (K_1)_{12}\,(K_1)_{23}}{(K_1)_{23}^2 -(K_1)_{22}\,(K_1)_{33}} & -\frac{(K_1)_{23}}{(K_1)_{33}} & 1
\end{array}
\right)\,,
\end{equation}
and $\det(R_1)=1$. We stress that $R_1$ is not orthogonal. The transformed action takes the form

\begin{equation}
I^{(2)}_{\rm even} = \frac{M_p^2}{2}\,\int N\,dt\,dk_L\,d^2k_T\,a^3 \left[\frac{\dot{{\cal Y}_2}^\dagger}{N} \,K_2\, \frac{\dot{{\cal Y}}_2}{N} +\frac{\dot{{\cal Y}}_2^\dagger}{N}\,M_2\,{\cal Y}_2+{\cal Y}_2^\dagger\,M_2^T\,\frac{\dot{{\cal Y}}_2}{N}
-{\cal Y}_2^\dagger\, \Omega^2_2\,{\cal Y}_2
\right]\,,
\end{equation}
where the new matrices are given by
\begin{eqnarray}
K_2 &=& R_1^T\,K_1\,R_1 \,,\nonumber\\
M_2 &=& R_1^T\,M_1\,R_1 + R_1^T\,K_1\,\frac{\dot{R}_1}{N}\,,\nonumber\\
\Omega_2^2 &=& R_1^T\,\Omega_1^2\,R_1 - \frac{\dot{R}_1^T}{N}\,M_1\,R_1 - R_1^T\,M_1^T\,\frac{\dot{R}_1}{N} - \frac{\dot{R}^T_1}{N}\,K_1\,\frac{\dot{R}_1}{N}\,.
\end{eqnarray}
For the vacuum configuration with $\Lambda\neq0$, the order of time derives can be determined from the background equations in Sec.\ref{sec:background} as
\begin{equation}
\dot{H} = {\cal O}(\epsilon^2)\,,\qquad
\dot{p}_L = -H\,N\,p_L + {\cal O}(\epsilon)\,,\qquad
\dot{p}_T = -H\,N\,p_T + {\cal O}(\epsilon)\,,\qquad
\dot{M}_{GW} = -M_{GW}\,\frac{\dot{r}}{2\,(1-r)} + {\cal O}(\epsilon)\,.
\end{equation}
Using these, we can obtain the components of the new matrices as:
\begin{eqnarray}
\left(K_2\right)_{11} &=& \frac{p_T^4}{8\,p^4}+{\cal O}(\epsilon)\,,\nonumber\\
\left(K_2\right)_{22} &=& -\frac{2\,a^4\,M_{GW}^2\,p_L^2}{1-r^2}\,\sigma + {\cal O}(\epsilon^2)\,,\nonumber\\
\left(K_2\right)_{33} &=& \frac{a^4\,M_{GW}^2\,p_T^2}{1-r^2}\,\sigma + {\cal O}(\epsilon^2)\,,
\end{eqnarray}
\begin{eqnarray}
\left(M_2\right)_{11} &=& \frac{p_T^4}{8\,H\,p^2}+{\cal O}(\epsilon)\,,\nonumber\\
\left(M_2\right)_{12} &=& -\frac{a^2\,M_{GW}^2\,p_L^2\,p_T^2\,\left(2\,p^2+p_T^2\,r\right)}{4\,H\,p^4\,(1-r)}\,\sigma +{\cal O}(\epsilon^2)
\,,\nonumber\\
\left(M_2\right)_{13} &=& \frac{a^2\,M_{GW}^2\,p_T^4\,\left(p^2-p_T^2\,r\right)}{4\,H\,p^4\,(1-r)}\,\sigma +{\cal O}(\epsilon^2)
\,,\nonumber\\
\left(M_2\right)_{21} &=& -\frac{a^2\,M_{GW}^2\,p_L^2\,p_T^2}{2\,H\,p^2\,(1-r)}\,\sigma +{\cal O}(\epsilon^2)
\,,\nonumber\\
\left(M_2\right)_{22} &=& -\frac{2\,a^4\,M_{GW}^2\,p_L^4\,r}{3\,H\,(1-r^2)}\,\sigma +{\cal O}(\epsilon^2)
\,,\nonumber\\
\left(M_2\right)_{23} &=& -\frac{2\,a^4\,M_{GW}^2\,p_L^2\,p_T^2\,r}{3\,H\,(1-r^2)}\,\sigma +{\cal O}(\epsilon^2)
\,,\nonumber\\
\left(M_2\right)_{31} &=& \frac{a^2\,M_{GW}^2\,p_T^4}{4\,H\,p^2\,(1-r)}\,\sigma +{\cal O}(\epsilon^2)
\,,\nonumber\\
\left(M_2\right)_{32} &=& \frac{a^4\,M_{GW}^2\,p_L^2\,p_T^2\,r}{3\,H\,(1-r^2)}\,\sigma +{\cal O}(\epsilon^2)
\,,\nonumber\\
\left(M_2\right)_{33} &=& \frac{a^4\,M_{GW}^2\,p_T^4\,r}{3\,H\,(1-r^2)}\,\sigma +{\cal O}(\epsilon^2)
\,,
\end{eqnarray}
\begin{eqnarray}
\left(\Omega_2\right)_{11} &=& \frac{M_{GW}^2\,p^4}{8\,p^4} + {\cal O}(\epsilon)\,,\nonumber\\
\left(\Omega_2\right)_{12} &=& {\cal O}(\epsilon)\,,\nonumber\\
\left(\Omega_2\right)_{13} &=& {\cal O}(\epsilon)\,,\nonumber\\
\left(\Omega_2\right)_{22} &=& \frac{1}{6}\,a^4\,M_{GW}^2\,p_L^2\,\left(3\,p^2+p_L^2\right) + {\cal O}(\epsilon)\,,\nonumber\\
\left(\Omega_2\right)_{23} &=& \frac{1}{6}\,a^4\,M_{GW}^2\,p_L^2\,p_T^2\,+ {\cal O}(\epsilon)\,,\nonumber\\
\left(\Omega_2\right)_{33} &=& \frac{1}{6}\,a^4\,M_{GW}^2\,p_T^2\,\left(3\,p^2+p_T^2\right) + {\cal O}(\epsilon)\,,\nonumber\\
\end{eqnarray}
\subsection{Normalizing the kinetic matrix}
We now do the following field rescaling,
\begin{equation}
{\cal Y}_3 = R_2^{-1}\,{\cal Y}_2\,,
\end{equation}
where
\begin{equation}
R_2 = {\rm diag}\,\left[2\,\sqrt{2}\,\frac{p^2}{p_T^2} \;\,,\;
\sqrt{\frac{1-r^2}{2\,\sigma}}\,\frac{1}{a^2\,M_{GW}\,p_L}\;\,,\;
\sqrt{\frac{1-r^2}{\sigma}}\,\frac{1}{a^2\,M_{GW}\,p_T} \right]\,,
\end{equation}
where we assumed $(1-r)\sigma>0$. 
With this transformation, the action becomes
\begin{equation}
I^{(2)}_{\rm even} = \frac{M_p^2}{2}\,\int N\,dt\,dk_L\,d^2k_T\,a^3 \left[\frac{\dot{{\cal Y}_3}^\dagger}{N} \,K_3\, \frac{\dot{{\cal Y}}_3}{N} +\frac{\dot{{\cal Y}}_3^\dagger}{N}\,M_3\,{\cal Y}_3+{\cal Y}_3^\dagger\,M_3^T\,\frac{\dot{{\cal Y}}_3}{N}
-{\cal Y}_3^\dagger\, \Omega^2_3\,{\cal Y}_3
\right]\,,
\end{equation}
where the new matrices are
\begin{eqnarray}
K_3 &=& R_2^T\,K_2\,R_2 \,,\nonumber\\
M_3 &=& R_2^T\,M_2\,R_2 + R_2^T\,K_2\,\frac{\dot{R}_2}{N}\,,\nonumber\\
\Omega_3^2 &=& R_2^T\,\Omega_2^2\,R_2 - \frac{\dot{R}_2^T}{N}\,M_2\,R_2 - R_2^T\,M_2^T\,\frac{\dot{R}_2}{N} - \frac{\dot{R}^T_2}{N}\,K_2\,\frac{\dot{R}_2}{N}\,,
\end{eqnarray}
with components,
\begin{equation}
K_3 = \left(
\begin{array}{ccc}
1+{\cal O}(\epsilon) & {\cal O}(\epsilon^{3/2})& {\cal O}(\epsilon^{3/2})\\
{\cal O}(\epsilon^{3/2})& -1 + {\cal O}(\epsilon)& {\cal O}(\epsilon^{3/2})\\
{\cal O}(\epsilon^{3/2}) & {\cal O}(\epsilon^{3/2}) & 1+{\cal O}(\epsilon) 
\end{array}
\right)\,,
\end{equation}
\begin{eqnarray}
\left(M_3\right)_{11} &=& \frac{p^2}{H} + {\cal O}(\epsilon)\,,\nonumber\\
\left(M_3\right)_{12} = \left(M_3\right)_{13} = \left(M_3\right)_{21} =\left(M_3\right)_{31} &=& {\cal O}(\sqrt{\epsilon})\,,\nonumber\\
\left(M_3\right)_{22} &=& H - \frac{p_L^2\,r}{3\,H}+\frac{\Sigma}{2\,\sigma}- \frac{\dot{r}}{N\,(1+r)}+{\cal O}(\epsilon)\,,\nonumber\\
\left(M_3\right)_{23} =-\left(M_3\right)_{32} &=& -\frac{\sqrt{2}\,p_L\,p_T\,r}{3\,H} + {\cal O}(\epsilon)\,,\nonumber\\
\left(M_3\right)_{33} &=& -H + \frac{p_T^2\,r}{3\,H} -\frac{\Sigma}{2\,\sigma}+ \frac{\dot{r}}{N\,(1+r)} +{\cal O}(\epsilon)\,,\nonumber\\
\end{eqnarray}
\begin{eqnarray}
\left(\Omega_3^2\right)_{11} &=& M_{GW}^2 +{\cal O}(\epsilon)\,,\nonumber\\
\left(\Omega_3^2\right)_{12} = \left(\Omega_3^2\right)_{13} &=& {\cal O}(\sqrt{\epsilon})\,,\nonumber\\
\left(\Omega_3^2\right)_{22} &=& \frac{(1-r^2)\,(3\,p^2 +p_L^2)}{12\,\sigma} + {\cal O}(\epsilon^0)\,,\nonumber\\
\left(\Omega_3^2\right)_{23} &=& \frac{(1-r^2)\,p_L\,p_T}{6\,\sqrt{2}\,\sigma} + {\cal O}(\epsilon^0)\,,\nonumber\\
\left(\Omega_3^2\right)_{33} &=& \frac{(1-r^2)\,(3\,p^2 +p_T^2)}{6\,\sigma} + {\cal O}(\epsilon^0)\,.
\end{eqnarray}
\subsection{Anti-symmetrizing the mixing matrix}
We now add the boundary term:
\begin{equation}
I^{(2)}_{\rm even} \to I^{(2)}_{\rm even} - \frac{M_p^2}{4}\,\int N\,dt\,dk_L\,d^2k_T\,\frac{1}{N}\frac{d}{dt}\left[a^3 {\cal Y}_3^\dagger\,\left(M_3+M_3^T\right)\,{\cal Y}_3\right]\,,
\end{equation}
giving the new action, 
\begin{equation}
I^{(2)}_{\rm even} = \frac{M_p^2}{2}\,\int N\,dt\,dk_L\,d^2k_T\,a^3 \left[\frac{\dot{{\cal Y}_3}^\dagger}{N} \,K_3\, \frac{\dot{{\cal Y}}_3}{N} +\frac{\dot{{\cal Y}}_3^\dagger}{N}\,M_4\,{\cal Y}_3-{\cal Y}_3^\dagger\,M_4\,\frac{\dot{{\cal Y}}_3}{N}
-{\cal Y}_3^\dagger\, \Omega^2_4\,{\cal Y}_3
\right]\,,
\end{equation}
where the mixing matrix is now antisymmetric
\begin{equation}
M_4 = \frac{1}{2}\,\left(M_3 - M_3^T\right) = 
\left(
\begin{array}{ccc}
{\cal O}(\epsilon) & {\cal O}(\sqrt{\epsilon}) & {\cal O}(\sqrt{\epsilon}) \\
{\cal O}(\sqrt{\epsilon})& {\cal O}(\epsilon) & -\frac{p_L\,p_T\,r}{2\,\sqrt{2}\,H}+{\cal O}(\epsilon)\\
{\cal O}(\sqrt{\epsilon})& \frac{p_L\,p_T\,r}{2\,\sqrt{2}\,H}+{\cal O}(\epsilon) & {\cal O}(\epsilon)\,,
\end{array}
\right)
\end{equation}
and the frequency matrix is
\begin{equation}
\Omega_4^2 = \Omega_3^2 +\frac{1}{2\,N\,a^3}\,\frac{d}{dt}\,\left[a^3\,\left(M_3+M_3^T\right)\right]=
\left(
\begin{array}{ccc}
p^2 + M_{GW}^2 + {\cal O}(\epsilon) & {\cal O}(\sqrt{\epsilon}) & {\cal O}(\sqrt{\epsilon}) \\
{\cal O}(\sqrt{\epsilon})&
\frac{(1-r^2)\,(3\,p^2+p_L^2)}{12\,\sigma} +  {\cal O}(\epsilon^0) &
\frac{(1-r^2)\,p_L\,p_T}{6\,\sqrt{2}\,\sigma} +  {\cal O}(\epsilon^0) \\
{\cal O}(\sqrt{\epsilon}) & 
\frac{(1-r^2)\,p_L\,p_T}{6\,\sqrt{2}\,\sigma} +  {\cal O}(\epsilon^0) &
\frac{(1-r^2)\,(3\,p^2+p_T^2)}{6\,\sigma} +  {\cal O}(\epsilon^0)
\end{array}
\right)\,.
\end{equation}
At this point, we see that the first mode is decoupled at the leading order, and has the same dispersion relation as GW in FLRW,
\begin{equation}
\omega_1^2 = p^2 +M_{GW}^2 + {\cal O}(\epsilon)\,.
\end{equation}
From here on, we concentrate on the remaining two degrees only. 

\subsection{Removing the mixing}
The action for the two degrees is now,
\begin{equation}
I^{(2)}_{\rm even,2,3} = \frac{M_p^2}{2}\,\int N\,dt\,dk_L\,d^2k_T\,a^3 \left[\frac{\dot{{\cal Y}_5}^\dagger}{N} \,K_5\, \frac{\dot{{\cal Y}}_5}{N} +\frac{\dot{{\cal Y}}_5^\dagger}{N}\,M_5\,{\cal Y}_5-{\cal Y}_5^\dagger\,M_5\,\frac{\dot{{\cal Y}}_5}{N}
-{\cal Y}_5^\dagger\, \Omega^2_5\,{\cal Y}_5
\right]\,,
\end{equation}
where ${\cal Y}_5 = \left[\,({\cal Y}_3)_2 \;\,,\; ({\cal Y}_3)_3  \, \right]$ and
\begin{equation}
K_5 = \left(
\begin{array}{cc}
-1 + {\cal O}(\epsilon)& {\cal O}(\epsilon^{3/2})\\
{\cal O}(\epsilon^{3/2}) & 1+{\cal O}(\epsilon) 
\end{array}
\right)\,,
\end{equation}
\begin{equation}
M_5 = 
\frac{p_L\,p_T\,r}{2\,\sqrt{2}\,H}\,\left(
\begin{array}{ccc}
0& -1\\
1& 0\,,
\end{array}
\right) + {\cal O}(\epsilon)
\end{equation}
\begin{equation}
\Omega_5^2 = \frac{(1-r^2)}{12\,\sigma}\,\left(
\begin{array}{cc}
(3\,p^2+p_L^2)&
\sqrt{2}\,p_L\,p_T\\
\sqrt{2}\,p_L\,p_T&
2\,(3\,p^2+p_T^2)
\end{array}
\right) +  {\cal O}(\epsilon^0)\,.
\end{equation}
The next field transformation is
\begin{equation}
 {\cal Y}_6 = R_5^{-1}\,{\cal Y}_5\,,
\end{equation}
where
\begin{equation}
R_5 = \left(
\begin{array}{cc}
\cosh[\theta(t)] &\sinh[\theta(t)] \\
\sinh[\theta(t)] &\cosh[\theta(t)] 
\end{array}
\right)\,,
\end{equation}
where the transformation function satisfies
\begin{equation}
\frac{\dot{\theta}}{N} = -\frac{p_L\,p_T\,r}{2\,\sqrt{2}\,H}\,.
\end{equation}
In the new basis, the mixing matrix becomes 
\begin{equation}
M_6 = R_5^T\,M_5\,R_5 + R_5^T\,K_5 \, \frac{\dot{R}^T_5}{N}  = {\cal O}(\sqrt{\epsilon})\,,
\end{equation}
whereas the frequency matrix reads
\begin{equation}
\Omega_6^2 = R_5^T\,\Omega_5^2\,R_5 - \frac{\dot{R}_5^T}{N}\,M_5\,R_5 - R_5^T\,M_5^T\,\frac{\dot{R}_5}{N} - \frac{\dot{R}^T_5}{N}\,K_5\,\frac{\dot{R}_5}{N} = R_5^T\,\Omega_5^2\,R_5 + {\cal O}(\epsilon^0)\,,
\end{equation}

Finally, the action is, at the relevant order
\begin{equation}
I^{(2)}_{\rm even,2,3} = \frac{M_p^2}{2}\,\int N\,dt\,dk_L\,d^2k_T\,a^3 \left[\frac{\dot{{\cal Y}_6}^\dagger}{N} \,K_5\, \frac{\dot{{\cal Y}}_6}{N} -{\cal Y}_6^\dagger\, (R_5^T\,\Omega^2_5\,R_5)\,{\cal Y}_6
\right]\,.
\end{equation}

\subsection{Eigenfrequencies}
At this point, to diagonalize the frequency matrix, we still need time dependent transformations, which will inevitably reintroduce the mixing. In that sense, it does not seem possible to diagonalize the system at the level of the Lagrangian. This is very similar to the coupled bosons in external background, where the diagonalization is done at the Hamiltonian level \cite{Nilles:2001fg}. The extended calculation to include ghost degrees is presented in Appendix \ref{sec:hamiltonian}, where the eigenvalues of the frequency matrix turn out to be the same as the eigenfrequencies in the diagonalized Hamiltonian.

Thus, the eigenvalues of $\Omega_6$ will actually give us the correct dispersion relations. We diagonalize the matrix $\Omega_6$ through,
\begin{equation}
R_6^T\,\Omega_6 \,R_6 = R_6^T\,R_5^T\,\Omega_5 \,R_5\,R_6 = {\rm diag}\left(-\omega_2^2\;\,,\;\omega_3^2\right)\,,
\end{equation}
where
\begin{equation}
R_6 = \left(
\begin{array}{cc}
\cosh[\varphi(t)] &\sinh[\varphi(t)] \\
\sinh[\varphi(t)] &\cosh[\varphi(t)] 
\end{array}
\right)\,,
\end{equation}
and $\varphi$ satisfies
\begin{equation}
\cosh\left[2\,(\theta +\varphi)\right] =\frac{10\,p^2+p_T^2}{\sqrt{\left(10\,p^2+p_T^2\right)^2-8\,p_L^2\,p_T^2}}\,,
\qquad
\sinh\left[2\,(\theta +\varphi)\right] =-\frac{2\,\sqrt{2}\,p_L\,p_T}{\sqrt{\left(10\,p^2+p_T^2\right)^2-8\,p_L^2\,p_T^2}}\,.
\end{equation}
As a result, the eigenfrequencies are found to be
\footnote{We remind that in this calculation, we assumed that the second mode is a ghost, i.e. $(1-r)\,\sigma>0$. Assuming that the third mode is a ghost $(1-r)\,\sigma < 0$, one obtains similar expressions for the eigenfrequencies:
\begin{eqnarray}
\omega_2^2 &=& \left(\frac{r^2-1}{24\,\sigma}\right)\,\left[\sqrt{\left(10\,p^2+p_T^2\right)^2 - 8\,p_L^2\,p_T^2} - \left(2\,p^2 +3\,p_T^2\right) \right]\,,\nonumber\\
\omega_3^2 &=& -\left(\frac{r^2-1}{24\,\sigma}\right)\,\left[\sqrt{\left(10\,p^2+p_T^2\right)^2 - 8\,p_L^2\,p_T^2} + \left(2\,p^2 +3\,p_T^2\right) \right]\,.
\end{eqnarray}
}
\begin{eqnarray}
\omega_2^2 &=& -\left(\frac{1-r^2}{24\,\sigma}\right)\,\left[\sqrt{\left(10\,p^2+p_T^2\right)^2 - 8\,p_L^2\,p_T^2} - \left(2\,p^2 +3\,p_T^2\right) \right]\,,\nonumber\\
\omega_3^2 &=& \left(\frac{1-r^2}{24\,\sigma}\right)\,\left[\sqrt{\left(10\,p^2+p_T^2\right)^2 - 8\,p_L^2\,p_T^2} + \left(2\,p^2 +3\,p_T^2\right) \right]\,.
\end{eqnarray}

\section{Dispersion relations of coupled system with ghosts, in external background}
\label{sec:hamiltonian}
The action of the modes in Section \ref{sec:evenexpand} contains a time dependent and non-diagonal frequency matrix, and one of the degrees has a negative kinetic term. At the level of the Lagrangian, it is not possible to recover a diagonal form. In this Appendix, we extend the formalism of \cite{Nilles:2001fg} to include ghost degrees, show that the Hamiltonian can be diagonalized and obtain dispersion relations of each degree of freedom.

We start by an action of the form
\begin{equation}
I = \int\,d^3k\,dt\,{\cal L}_k=\frac{1}{2}\,\int d^3k\,dt\,\Big[\dot{\phi}^\dagger(t,{\bf k})\,\eta\,\dot{\phi}(t,{\bf k}) - \phi^\dagger(t,{\bf k})\,\Omega^2(t,{\bf k})\,\phi(t,{\bf k})\Big]\,,
\label{eq:action1}
\end{equation}
where $\phi$  is a $N=N_1 + N_2$ dimensional array of fields, with $N_1$ ghosts and $N_2$ physical fields. $\Omega^2$ is a $N\times N$ real and symmetric matrix, which stays invariant under ${\bf k} \to -{\bf k}$. The signature of the kinetic matrix is
\begin{equation}
\eta_{ij} = \delta_{ij}\,\left\{
\begin{array}{ll}
-1&\;\,, i,j \le N_1\\
+1&\;\,, i,j > N_1
\end{array}
\right.\,.
\end{equation}
The conjugate momentum array can be immediately found as
\begin{equation}
\pi_i \equiv \frac{\partial\,{\cal L}_k}{\partial\,\dot{\phi}^\dagger_i} = \eta_{ij}\,\dot{\phi}_j\,,
\qquad
\pi_i^\dagger \equiv \frac{\partial\,{\cal L}_k}{\partial\,\dot{\phi}_i} = \dot{\phi}_j^\dagger\,\eta_{ji}\,,
\end{equation}
giving the Hamiltonian
\begin{equation}
H = \int d^3k \,{\cal H}_k = \frac{1}{2}\,\int d^3k\,\Big[\pi^\dagger(t,{\bf k})\,\eta\,\pi(t,{\bf k})+\phi^\dagger(t,{\bf k})\,\Omega^2(t,{\bf k})\,\phi(t,{\bf k})\Big]\,.
\end{equation}
The Hamiltonian equations of motion are:
\begin{equation}
\dot{\phi}= \eta\,\pi\,,
\qquad
\dot{\pi}= - \Omega^2\,\phi\,.
\label{eq:hameqmom}
\end{equation}
We also define the matrix $C=C(t,{\bf k})$ which diagonalizes $\Omega^2$ through
\begin{equation}
C^T\,\Omega^2 \, C = \eta\,\omega^2\quad ({\rm diagonal})\,,
\end{equation}
where $C$ is an element of group $SO(N_2,N_1)$, i.e.
\begin{equation}
C^T\,\eta\,C = \eta\,.
\end{equation}
We also note that $C$ is invariant under ${\bf k}\to - {\bf k}$.

It turns out to be useful to include the matrix $C$ in the decomposition of the fields $\phi$ and momenta $\pi$ into mode functions in a basis of $N$ dimensional creation/annihilation operator arrays $\hat{a}$ and $\hat{a}^\dagger$ as
\begin{eqnarray}
\phi(t,{\bf k}) &=& C(t,{\bf k})\left[h(t,{\bf k})\,\hat{a}_{\bf k} + h^\star(t,{\bf k})\,\hat{a}^\dagger_{-\bf k}\right] \,,\nonumber\\
\pi(t,{\bf k}) &=& \left[C^T(t,{\bf k})\right]^{-1}\,\left[\tilde{h}(t,{\bf k})\,\hat{a}_{\bf k} + \tilde{h}^\star(t,{\bf k})\,\hat{a}^\dagger_{-\bf k}\right] \,,
\label{eq:decompphipi}
\end{eqnarray}
where mode functions $h$ and their conjugates $\tilde{h}$ are $N\times N$ matrices. With this decomposition, we can rewrite Eqs.(\ref{eq:hameqmom}) as
\begin{equation}
\dot{h} = \eta\,\tilde{h} - \Gamma\,h\,,\qquad
\dot{\tilde{h}} = -\eta\,\omega^2\,h+\Gamma^T\,\tilde{h}\,,
\end{equation}
where $\Gamma = C^{-1}\,\dot{C}$ corresponds to the rate of change of the transformation $C$.

Using the decomposition (\ref{eq:decompphipi}), the Hamiltonian density becomes
\begin{equation}
{\cal H}_k = \frac{1}{2}\,\left( \hat{a}_{\bf k}^\dagger \;\,,\; \hat{a}_{-{\bf k}}\right)\,
\left(
\begin{array}{ccc}
\tilde{h}^\dagger\,\eta\,\tilde{h} + h^\dagger \,\eta\,\omega^2\,h  
&& \tilde{h}^\dagger\,\eta\,\tilde{h}^\star + h^\dagger\,\eta\,\omega^2\,h^\star\\
\tilde{h}^T \,\eta\,\tilde{h}+ h^T\,\eta\,\omega^2\,h 
&& \tilde{h}^T\,\eta\,\tilde{h}^\star + h^T \,\eta\,\omega^2\,h^\star  
\end{array}
\right)\,
\left(
\begin{array}{l}
\hat{a}_{\bf k}\\
\hat{a}^\dagger_{-{\bf k}}
\end{array}
\right)\,.
\label{eq:hamden}
\end{equation}

Finally, we do a further redefinition, based on the solutions to the equations of motion in adiabatic regime,
\begin{equation}
h = \frac{1}{\sqrt{2\,\omega}}\,\left(\alpha+\beta\right)\,,
\qquad
\tilde{h} = - i\,\sqrt{\frac{\omega}{2}}\,\eta\,\left(\alpha-\beta\right)\,,
\label{eq:alphabeta}
\end{equation}
where $\alpha$ and $\beta$ are $N\times N$ complex matrices and are generalizations of Bogolyubov coefficients. We stress that this redefinition keeps generality.

Using the definition (\ref{eq:alphabeta}) in the Hamiltonian density (\ref{eq:hamden}), we obtain the Hamiltonian density
\begin{equation}
{\cal H}_k = \frac{1}{2}\,\left( \hat{b}_{\bf k}^\dagger \;\,,\; \hat{b}_{-{\bf k}}\right)\,
\left(
\begin{array}{cc}
\eta\,\omega &0\\
0& \eta\,\omega
\end{array}
\right)\,
\left(
\begin{array}{l}
\hat{b}_{\bf k}\\
\hat{b}^\dagger_{-{\bf k}}
\end{array}
\right)\,,
\end{equation}
where the new creation/annihilation operators are defined as
\begin{equation}
\left(\begin{array}{l}
\hat{b}_{\bf k}\\
\hat{b}^\dagger_{-{\bf k}}
      \end{array}
\right)
\equiv
\left(
\begin{array}{cc}
\alpha&\beta^\star\\
\beta& \alpha^\star
\end{array}
\right)\,
\left(\begin{array}{l}
\hat{a}_{\bf k}\\
\hat{a}^\dagger_{-{\bf k}}
      \end{array}
\right)\,.
\end{equation}
Since $\eta\,\omega$ is already diagonal, the Hamiltonian in the new basis is also diagonal. After normal ordering, we end up with
\begin{eqnarray}
{\cal H}_k &=& \frac{1}{2}\sum_{i=1}^N\left(\eta\,\omega\right)_i\,\hat{b}^\dagger_i\,\hat{b}_i \,.\nonumber\\
&=& -\frac{1}{2}\sum_{i=1}^{N_1}\omega_i\,\hat{b}^\dagger_i\,\hat{b}_i+\frac{1}{2}\sum_{i=1}^{N_2}\omega_i\,\hat{b}^\dagger_i\,\hat{b}_i  
\end{eqnarray}

This calculation shows that the eigenvalues of the frequency matrix in Eq.(\ref{eq:action1}) actually correspond to the eigenfrequencies of the modes in the diagonal basis.

\section{The situation of perturbations \emph{on} the fixed point}

Since the even mode action in Sec.\ref{evenanis} is very bulky, it was not possible to obtain the dispersion relations for each degree of freedom, at least not for arbitrary momentum. Observing that the matrix $\bar{M}$ which mixes first derivatives with the fields is of order ${\cal O}(\epsilon^0)$, it is then a fair question to ask whether the kinetic term is modified when the action is brought to a diagonal form. 

However, this reasoning requires integrating out the field ${\cal Y}_3=\hat{E}_\pi$, a procedure which cannot be performed in our case as the field ${\cal Y}_3$ does not have any mass-gap. Let us consider this issue in more detail. The term $\mu^2$ in the Lagrangian, where ${\cal L}\ni-\mu^2{\cal Y}_3^2$, is surely of order ${\cal O}(\epsilon^0)$ but multiplied by a factor $p_T^4$. Instead, the kinetic term of this same field ${\cal Y}_3$ -- which tends to exactly vanish on the fixed point -- is multiplied only by a factor $p_T^2$ as shown in Eq.~(\ref{eq:kinvan}). Rescaling the field by ${\cal Y}_3\to\tilde{\cal Y}_3/p_T$, makes the ${\cal O}(\epsilon)$-kinetic term momentum-independent; however, its mass -- being proportional to $p_T^2$ -- still tends to vanish for small $p_T$. This is tantamount to saying that we cannot in fact describe the mass of the field in the small transverse-momentum limit, that is $p_T\to0$, because we still have that $\lim_{p_T\to0}(\mu^2/p_T^2)\to0$ and, consequently, we cannot 
integrate the mode ${\cal Y}_3$ any more in this limit.

Therefore, we are led to deduce that we cannot, in general, study the perturbations on the exact-fixed-point solution, as this would lead to the inconsistency of being able to integrate out a mode which actually cannot be integrated out -- at least for small $p_T$. On the other hand, it makes sense to study that background which is \emph{not} the fixed point, but close enough to it. We have followed this last procedure in our analysis in this paper.

This situation also implies that, in order to check whether there is a strong-coupling limit on the fixed point solution, we need to analyze the Lagrangian at higher orders (cubic, at least) in the perturbations, and check whether there is indeed any non-trivial strong contribution/back-reaction coming from these higher order terms.

\section{Even sector dispersion relations in the IR}
\label{app:dispIR}

Unfortunately, for the perturbation analysis around the anisotropic fixed point discussed in Sec.\ref{sec:stabilityGLM}, the problem is technically very involved, and calculation of the full dispersion relations is very difficult. However, it is possible to finalize the diagonalization in the IR regime. This approach allows us to determine whether the even mode with vanishing kinetic term on the fixed point suffers from strong coupling or not. Notice that we already have one mode (in the odd sector) which has vanishing kinetic term and has no mass gap.

Due to the direction dependence of the background, in the IR regime where the momentum dependence is removed, the dispersion relations still depend on the orientation of the momentum vector. We decompose the momenta as
\begin{equation}
p_L = p\,\xi\,,\qquad
p_T = p\,\sqrt{1-\xi^2}\,,
\end{equation}
and take the limit $p\to0$. Proceeding the same way as we did for the FLRW solution (see Appendix \ref{app:diagonalize} for details), the diagonalization in this limit reveals,
\begin{equation}
\omega_1 = {\cal O}(p^0)\,,\qquad
\omega_2 = {\cal O}(p^0)\,,\qquad
\left.\omega_3 \right\vert_{p\to0}=  0 \,,
\end{equation}
where modes labeled $1$ and $2$ correspond to degrees which have non vanishing kinetic terms on the anisotropic fixed point, while for mode $3$, the kinetic term is zero.

Combining our result from the 2d vector sector, we observe that:
\begin{itemize}
\item Even though by appropriate choice of parameters, we may avoid ghost degrees, for the two modes which have vanishing kinetic terms, there is no mass gap. Therefore, we expect that they are infinitely strongly coupled. 
\item The mode with ${\cal O}(1)$ kinetic term in the odd sector is massless on the fixed point, with sound speed $c_s=1$.
\item The two modes with ${\cal O}(1)$ kinetic term in the even sector have masses, but these depend on the orientation of the momentum vector. The gradient part of the dispersion relation was not calculated.
\end{itemize}

For the two massive modes in the even sector, the mass terms are still to bulky for presentation. We end this section by looking at two extreme examples for the momentum direction.

\subsection{$\xi\to 1$, momentum is in the $\hat{x}$ direction}
In this case, the momentum is aligned with the privileged direction $\hat{x}$. The masses for this situation is
\begin{equation}
m_1^2 \simeq -3\,M^2\,,\qquad
m_2^2 \simeq \frac{3\,M^2\,\tilde{M}^2\,(9\,H_0^2+\tilde{M}^2)}{9\,H_0^2\,(\tilde{M}^2-3\,M^2)-\tilde{M}^4}
\,.
\end{equation}
If the no-ghost condition (\ref{noghost1}) and fixed point stability condition (\ref{fixedpointstability}) are satisfied, both squared-masses are positive.

\subsection{$\xi\to 0$, momentum is perpendicular to the $\hat{x}$ direction}
In this case, the momentum is along the $y-z$ plane. The masses of the modes then become
\begin{eqnarray}
m_1^2 &\simeq& -\frac{M^2\,\tilde{M}^2\left[\tilde{M}^8+36\,H_0^2\,\tilde{M}^4\,(\tilde{M}^2-3\,M^2)+243\,H_0^4\,(\tilde{M}^2-3\,M^2)^2\right]}{M^2\,\tilde{M}^8+27\,H_0^4\,(\tilde{M}^2-3\,M^2)^2(3\,M^2+2\,\tilde{M}^2)+18\,H_0^2\,\tilde{M}^4\,(3\,M^4-4\,M^2\,\tilde{M}^2+\tilde{M}^4)}\xi^2\,,\nonumber\\
m_2^2 &\simeq& \frac{M^4\,\left[9\,H_0^2\,(\tilde{M}^2-3\,M^2)-\tilde{M}^4\right]}{4\,H_0^2\left(\tilde{M}^2-3\,M^2\right)^2} + \frac{54\,H_0^2\,M^2\,\tilde{M}^2}{9\,H_0^2\,(\tilde{M}^2-3\,M^2)-\tilde{M}^4}\,\left[
1-\frac{9\,H_0^2\,(\tilde{M}^2-3\,M^2)-\tilde{M}^4}{12\,H_0^2(\tilde{M}^2-3\,M^2)}
\right]
\,.
\end{eqnarray}
Again, once the conditions (\ref{noghost1}) and (\ref{fixedpointstability}) are imposed, both squared-masses are positive.

\end{document}